\documentclass[aps,prb,twocolumn,floatfix,10pt,superscriptaddress]{revtex4-1}
\usepackage[utf8]{inputenc}

\usepackage{natbib}
\usepackage{graphicx}
\usepackage{latexsym}
\usepackage{amssymb}
\usepackage{amsmath}
\usepackage{amsfonts}
\usepackage{gensymb}
\usepackage{bm}
\usepackage{hyperref}
\usepackage{braket}
\usepackage{physics}
\usepackage{bbold}
\usepackage[caption=false]{subfig}
\usepackage[dvipsnames]{xcolor}
\usepackage[normalem]{ulem}

\hypersetup{
  pdfauthor = {Taige Wang, Wilbur Shirley, Xie Chen},
  pdftitle = {Wang et al. - 2019 - Foliated fracton order in the Majorana checkerboard model}
}

\begin{document}

\title{Foliated fracton order in the Majorana checkerboard model}

\author{Taige Wang}
\affiliation{Department of Physics and Institute for Quantum Information and Matter, California Institute of Technology, Pasadena, California 91125, USA}
\affiliation{Department of Physics, University of California, San Diego, California 92093, USA}

\author{Wilbur Shirley}
\author{Xie Chen}
\affiliation{Department of Physics and Institute for Quantum Information and Matter, California Institute of Technology, Pasadena, California 91125, USA}

\date{\today}

\begin{abstract}

We establish the presence of foliated fracton order in the Majorana checkerboard model. In particular, we describe an entanglement renormalization group transformation which utilizes toric code layers as resources of entanglement, and furthermore discuss entanglement signatures and fractional excitations of the model. In fact, we give an exact local unitary equivalence between the Majorana checkerboard model and the semionic X-cube model augmented with decoupled fermionic modes. This mapping demonstrates that the model lies within the X-cube foliated fracton phase.

\end{abstract}


\maketitle

\section{Introduction}

Gapped quantum systems, such as discrete gauge theories and fractional quantum Hall states, can reside in non-trivial phases in the absence of symmetry if they are `topological'.\cite{WenRigid} Such systems have low-energy effective descriptions given by topological quantum field theory (TQFT).\cite{Schwarz2000,witten1988, Atiyah1988} However, a class of recently discovered three-dimensional gapped lattice models known as \textit{fracton} models belong to non-trivial phases but defy such a characterization.\cite{ChamonModel,ChamonModel2,HaahCode,YoshidaFractal,HaahRG,VijayFracton,Sagar16,MaLayers,Slagle2Spin,Slagle17Lattices,HsiehPartons,HalaszSpinChains,CageNet,SongTwistedFracton,VijayNonabelian,FractonRev,GeneralizedHaah,Lego,StringMembraneNet}
Their most salient, unifying properties are the presence of point-like fractional excitations with fundamentally constrained mobility and a degenerate ground space which grows exponentially with linear system size. These features preclude a TQFT description.

A particularly exotic class of fracton models are the fractal spin liquids (i.e. the Type-II models), in which the operators that transport point-like fractional excitations are constrained to have certain fractal geometries.\cite{HaahCode,YoshidaFractal,HaahRG} A somewhat more terrestrial class of models (the Type-I family) exhibit three categories of point-like excitations: \textit{fractons}, which are fully immobile, \textit{lineons}, which can move along a line, and \textit{planons}, which are mobile within a plane.\cite{ChamonModel2,VijayFracton} The concept of \textit{foliated fracton order} was introduced recently in an attempt to systematize the study of these Type-I models.\cite{3manifolds,FractonStatistics,FractonEntanglement,Gauging,Checkerboard} This notion builds on the observation that many of these models have a foliated structure of long-range entanglement, in the sense that layers of 2D topological orders can be disentangled from the bulk by local unitary operations. The identification of this structure has shed light on the scaling of ground space growth, the structure of fractional excitations in such models, and entanglement entropic signatures discussed previously in the literature.\cite{ShiEntropy,HermeleEntropy,BernevigEntropy,FractonEntanglement} Furthermore, a more coarse notion of gapped phases of matter is motivated by this observation: in particular, a \textit{foliated fracton phase} is defined as an equivalence class of Hamiltonians under adiabatic deformation augmented with the possible addition of layers of 2D topological orders.

It remains unclear to what extent this framework captures known Type-I models. Partial progress has been made toward understanding the phase relations between these models,\cite{Checkerboard,FractonStatistics} but the picture is far from complete. Moreover, all examples of foliated fracton order that have been studied thus far are in models with bosonic degrees of freedom, and it is not yet clear whether the notion can be extended to fermionic models. 

In this paper we address these questions by demonstrating that a prototypical example, the Majorana checkerboard model introduced in Ref. \onlinecite{VijayFracton}, exhibits foliated fracton order. In fact, we find that this model is actually a fermionic version of a previously known fractonic spin model called the semionic X-cube model, which was originally described via the coupled layers construction of Ref. \onlinecite{MaLayers}. As it has been shown that the semionic X-cube model has the same foliated fracton order as the X-cube model,\cite{FractonStatistics} the Majorana checkerboard model thus has the same order as well.

The paper's contents are as follows: in Sec. \ref{sec:model}, we briefly review the Majorana checkerboard model. In Sec. \ref{sec:rg}, we describe a renormalization group (RG) transformation for the model which utilizes layers of toric code as resources of entanglement, hence establishing its foliated fracton order. In Sec. \ref{sec:entanglement} we discuss entanglement entropic signatures of the foliated fracton order in the model, and in Sec. \ref{sec:excitation} we discuss the structure of quotient superselection sectors (QSS). In the following Sec. \ref{sec:mapping1}, we describe a mapping from the Majorana checkerboard model to a spin Hamiltonian (plus decoupled fermions), and in Sec. \ref{sec:mapping2} a mapping from this stabilizer code spin Hamiltonian to the semionic X-cube model, hence establishing its equivalence to the X-cube model as a foliated fracton order. Finally we conclude with a discussion in Sec. \ref{sec:discussion}.

\section{The Majorana Checkerboard Model}
\label{sec:model}

\begin{figure}[htbp]
    \centering
    \includegraphics[width=0.45\textwidth]{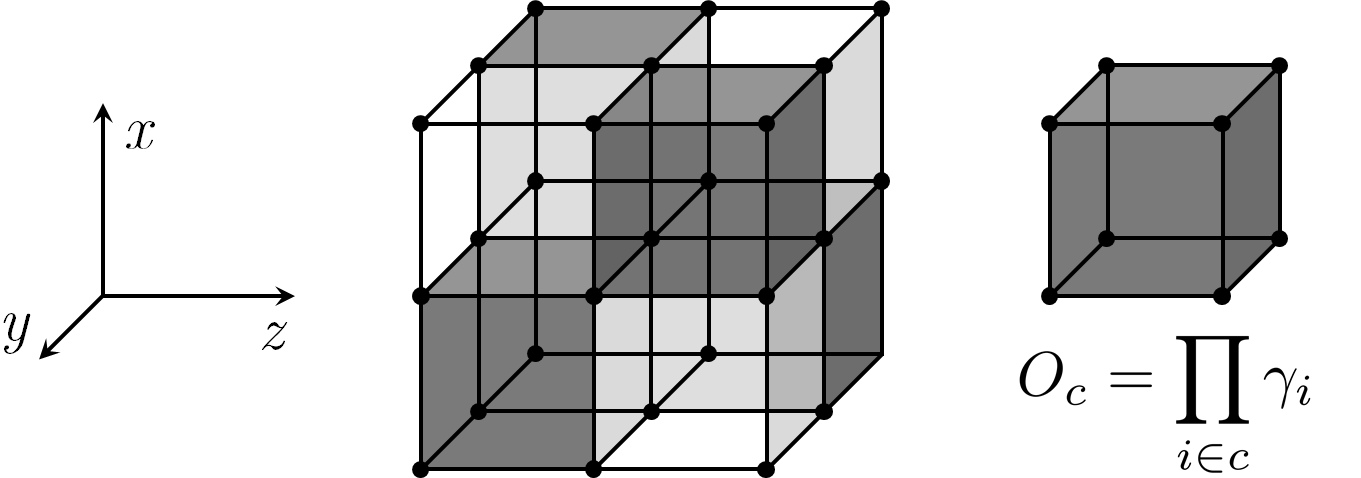}
    \caption{Bipartition of a cubic lattice into $A$ (shaded) and $B$ (unshaded) checkerboard sublattices. Majorana fermions are placed at the vertices of the lattice. The operator $O_c$ acts on cubes $c$ in the $A$ sublattice and is defined as the product of the 8 Majoranas at the corners of cube $c$.}
    \label{fig:checkerboardhamiltonian}
\end{figure}

The Majorana checkerboard model was first introduced in Ref. \onlinecite{VijayFracton} as a Majorana stabilizer code with one Majorana fermion on each vertex of a cubic lattice. The elementary cubes are bipartitioned into $A$-$B$ checkerboard sublattices (as shown in Fig. \ref{fig:checkerboardhamiltonian}), and the Hamiltonian is given by
\begin{equation}
    H = -\sum_{c \in A} O_c
\end{equation}
where $O_c=\prod_{i \in c} \gamma_i$ is the product of the eight Majorana operators at the corners of cube $c$. The Hamiltonian terms mutually commute as they share either zero or two Majorana operators, and their energies can be simultaneously minimized. The model exhibits a ground state degeneracy (GSD) on a $2L_x \times 2L_y \times 2L_z$ cubic lattice under periodic boundary conditions which satisfies\cite{VijayFracton}
\begin{equation}
    \log_2 \text{GSD} = 2L_x+2L_y+2L_z - 3.
\end{equation}
Note that the number of logical qubits in the ground space is half that of the spin checkerboard model on the same lattice\cite{Sagar16}, as per the doubling lemma of Ref. \onlinecite{MFC}.

\begin{figure}[htbp]
    \centering
    \includegraphics[width=0.4\textwidth]{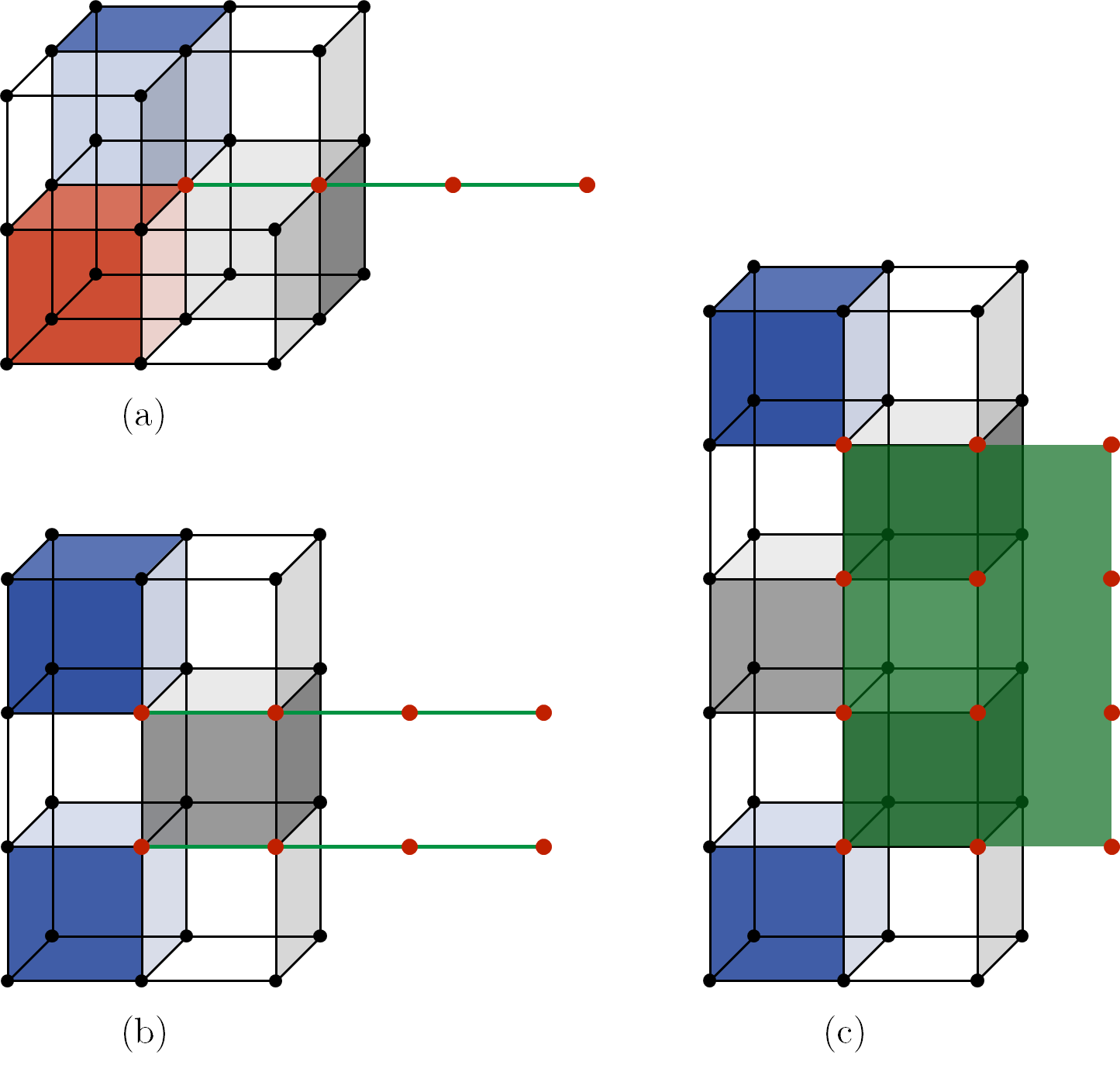}
    \caption{Point-like excitations in the Majorana checkerboard model. The colored cubes correspond to stabilizer terms which are violated by a given excitation. The operator which creates a given excitation is denoted by the product of the red Majoranas depicted. (a) A \textit{lineon} created at the end of a rigid string operator (green). (b) A \textit{planon} created at the end of a flexible string operator. (c) A \textit{fracton} created at the corner of a rectangular membrane operator (green).}
    \label{fig:majoranaexcitation}
\end{figure}

As discussed in detail in Ref. \onlinecite{VijayFracton}, the model exhibits point-like excitations with a dimensional hierarchy of constrained mobility as depicted in Fig. \ref{fig:majoranaexcitation}. Fractons, which are fundamentally immobile, are created at the corners of rectangular membrane operators. Lineons, which can move along a line only, are created at the endpoints of rigid string operators and can be thought of as composites of two fractons. Finally, planons, which are free to move within a plane, can be thought of as composites of two lineons, or as composites of two fractons in their own right. In Sec. \ref{sec:excitation}, we discuss how the notion of \textit{quotient superselection sectors} can be used to analyze the fractional excitations of the model.

\section{Entanglement renormalization} \label{sec:rg}

\begin{figure}[htbp]
    \includegraphics[width=0.4\textwidth]{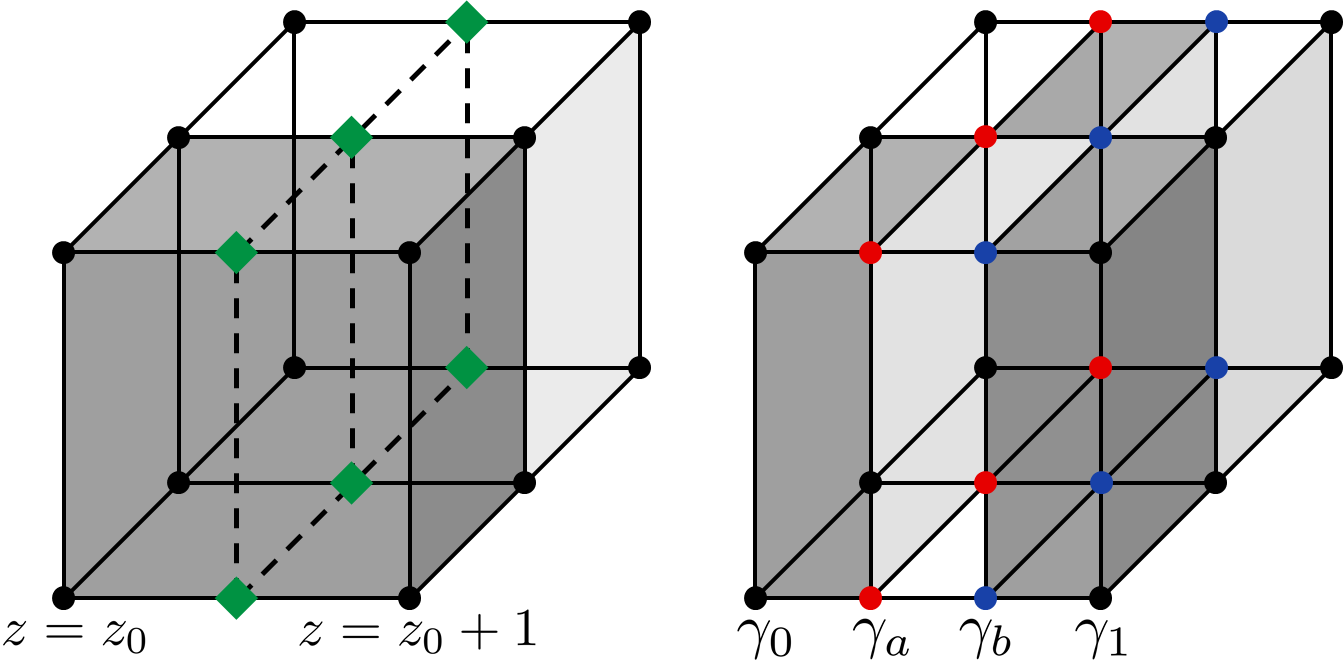}
    \caption{Degrees of freedom in (left) the original Majorana checkerboard model (black dots represent Majorana fermions) augmented with one copy of the toric code (green diamonds represent qubits), and (right) the enlarged Majorana checkerboard model, in which the red and blue dots represent added Majoranas along $z=a$ and $z=b$ and the black dots correspond to the original Majoranas.}
    \label{fig:rg}
\end{figure}

In this section, we discuss an entanglement renormalization group (RG) transformation\cite{VidalRG,ER2,Stringnet,Xie10} for the Majorana checkerboard model, which utilizes copies of the toric code as 2D resource layers and thus establishes the presence of foliated fracton order in the model. It can be compared to the analogous RG transformation for the X-cube model, which also utilizes toric code resource layers.\cite{3manifolds} The transformation consists of a fermion parity-preserving local unitary map $S$ between the Majorana checkerboard model on a $2L_x\times2L_y\times2L_z$ cubic lattice (described by Hamiltonian $H_0$), augmented with one copy of the toric code ($H_\mathrm{2D})$, and the Majorana checkerboard model on a $2L_x\times2L_y\times2(L_z+1)$ size lattice ($H_1$):
\begin{equation}
    S (H_0+H_\mathrm{2D})S^\dagger
    \cong H_1.
\end{equation}
Here the relation $\cong$ denotes that the two Hamiltonians are equivalent as stabilizer codes and thus have identical ground spaces. We call the 2D topological layers the ``resource layers" for the RG transformation. An equivalent transformation applies in the $x$ and $y$ directions as well.

\begin{figure}[htbp]
    \includegraphics[width=0.43\textwidth]{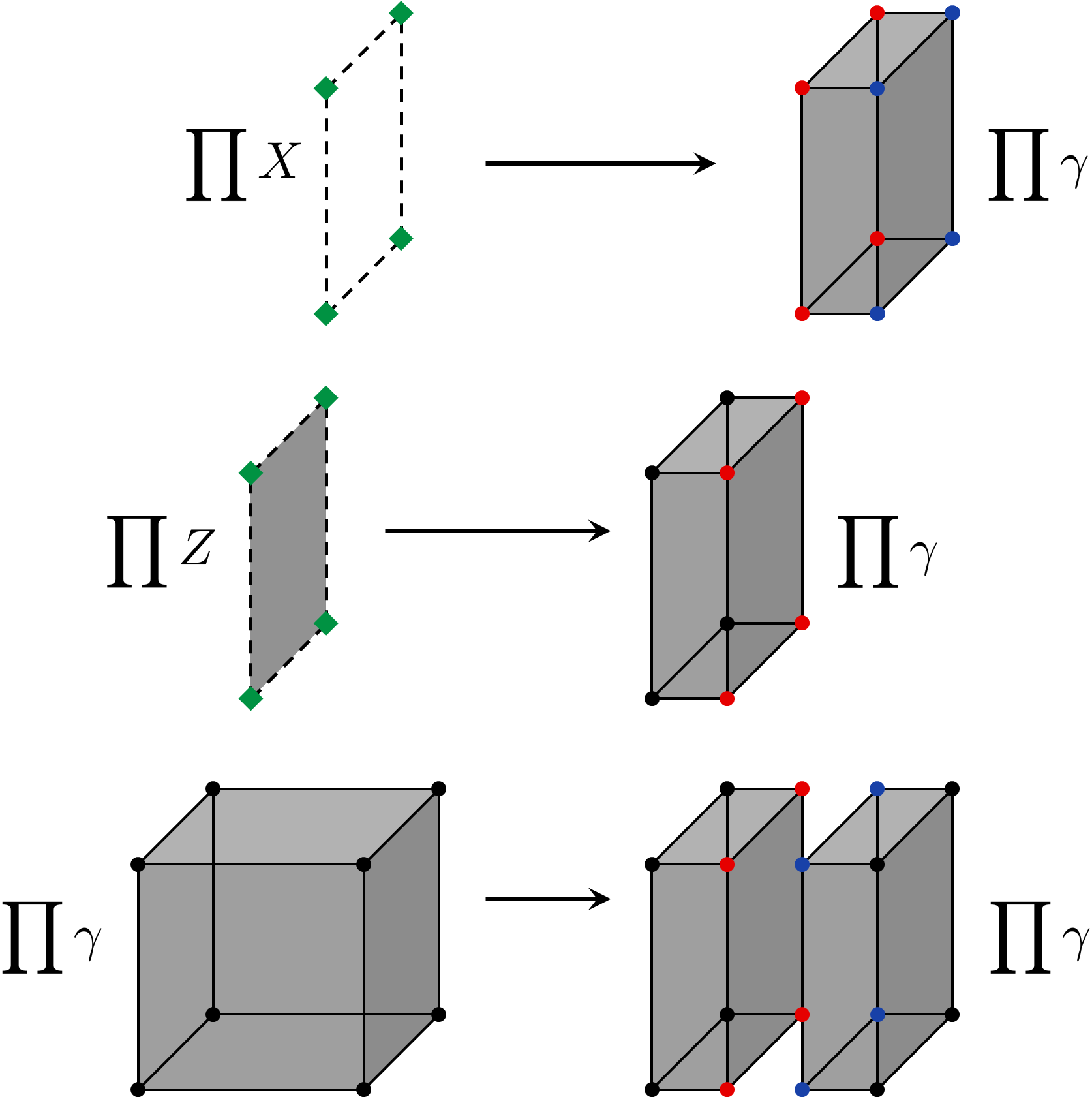}
    \caption{Mapping of Hamiltonian stabilizers under the local unitary transformation $S$.}
    \label{fig:rgmapping}
\end{figure}

In particular, suppose the toric code layer is inserted between layers $z_0$ and $z_0+1$ of the original lattice. Its degrees of freedom consist of qubits placed between the lattice sites of these two layers, as shown in Fig. \ref{fig:rg}. Its Hamiltonian is given as 
\begin{equation}
    H_{2D} = - \sum_{p \in A}\prod_{i\in p}Z_i- \sum_{p \in B}\prod_{i\in p}X_i.
\end{equation}
Here, the 2D $A$-$B$ checkerboard sublattices coincide with the 3D $A$-$B$ checkerboard sublattices. The unitary $S$ maps the combined Majorana and spin degrees of freedom to a pure Majorana system with two additional Majoranas on the links between $z_0$ and $z_0+1$. The latter system constitutes an enlarged $2L_x\times2L_y\times2(L_z+1)$ size cubic lattice of Majorana fermions. The two systems have identical Hilbert spaces. To see this, for each $(x,y)$ coordinate, denote the Majorana at $z=z_0$ by $\gamma_0$, the Majorana at $z=z_1$ by $\gamma_1$, and the added Majoranas by $\gamma_a$ and $\gamma_b$ (as in Fig. \ref{fig:rg}). On the left hand side of Fig. \ref{fig:rg}, the combination of $\gamma_0$, $\gamma_1$, and the spin forms a four-dimensional Hilbert space whose operator algebra is generated by $\gamma_0$, $\gamma_1$, $X$, and $Z$. On the right hand side of Fig. \ref{fig:rg}, the combination of $\gamma_0$, $\gamma_a$, $\gamma_b$, and $\gamma_1$ also forms a four-dimensional Hilbert space. The two sides can be mapped into each other under the following correspondence of operators:
\begin{equation}
    X \to \gamma_a\gamma_b, \quad Z \to \gamma_0\gamma_a, \quad \gamma_0 \to \gamma_0 \gamma_a\gamma_b, \quad \gamma_1\to\gamma_1.
\end{equation}
This mapping preserves the commutation relations of the local operator algebra at each $(x,y)$ coordinate as well as the global fermionic parity, hence it describes a parity-preserving local unitary transformation. In fact, it is exactly the local unitary map $S$ that is needed to implement the RG transformation.
Fig. \ref{fig:rgmapping} illustrates the mapping of Hamiltonian stabilizers under this unitary. Evidently, the resultant Hamiltonian generates the same stabilizer group as the enlarged Majorana checkerboard Hamiltonian. In other words, we find that the ground space of the original model tensored with the added toric code ground space is local unitarily equivalent to the ground space of the enlarged Majorana checkerboard model.

\section{Entanglement Signatures}
\label{sec:entanglement}

\begin{figure}[htbp]
    \centering
    \includegraphics[width=0.4\textwidth]{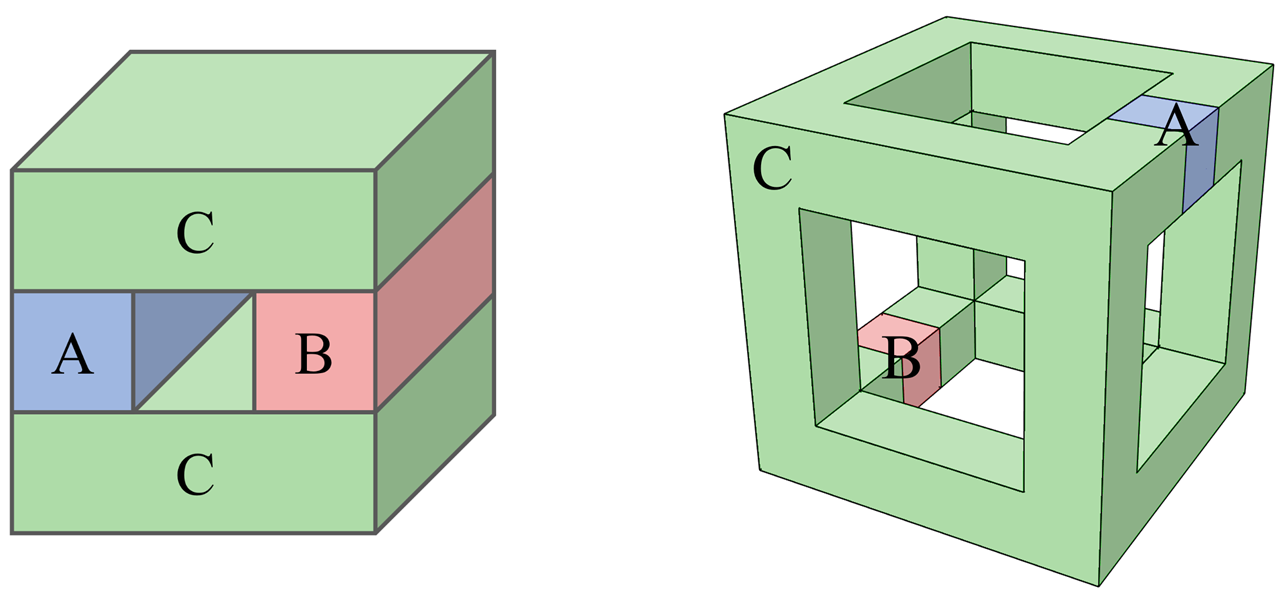}
    \caption{(Left) Solid torus and (right) wireframe entanglement entropy schemes.}
    \label{fig:entanglement}
\end{figure}

In this section, we briefly discuss the structure of entanglement entropy in the Majorana checkerboard model. Two entanglement schemes, solid torus and wireframe, among others, have proven useful in characterizing the order in foliated fracton models.\cite{ShiEntropy,HermeleEntropy,FractonEntanglement} In each scheme the quantity to be computed is the conditional mutual information
\begin{equation}
    I(A;B|C)=S_{AC}+S_{BC}-S_{C}-S_{ABC}
\end{equation}
where $S_R$ refers to the entanglement entropy of region $R$. The geometries of the $A$, $B$, and $C$ regions for the two schemes are depicted in Fig. \ref{fig:entanglement}. These schemes generalize the notion of topological entanglement entropy in two dimensions.\cite{EntropyKP,EntropyLW}

A simple technique for computing the ground state entanglement entropy of generic Majorana stabilizer codes is discussed in Appendix \ref{sec:majoranaentanglement}. Applied to the Majorana checkerboard model, one finds that $I(A;B|C)=2L+1$ for the solid torus scheme where $L$ is the length of the overall cubic region measured in twice the lattice constant, and $I(A;B|C)=1$ for the wireframe scheme. For both schemes these results hold provided the overall cubic region is aligned with the axes of the cubic lattice. (In fact, the entanglement entropy of the Majorana checkerboard model for a given region is exactly half of that for the equivalent region of the spin checkerboard model.\cite{FractonEntanglement})

As discussed in Ref. \onlinecite{FractonEntanglement}, the solid torus scheme serves as a diagnostic of the underlying foliation structure, and indeed the result is consistent with the triple foliation structure composed of 2D toric code layers identified in the RG transformation of the section prior. On the other hand, the wireframe scheme is engineered such that the contributions from the foliating layers completely cancel, resulting in a constant value which characterizes the foliated fracton phase. In the case of the Majorana checkerboard model, the result $I(A;B|C)=1$ is consistent with our finding that the model belongs to the X-cube foliated fracton phase, as discussed in Sections \ref{sec:mapping1} and \ref{sec:mapping2}.

\section{Quotient Superselection Sectors} \label{sec:excitation}

In Ref. \onlinecite{FractonStatistics}, the notion of \textit{quotient superselection sectors} was introduced as a way to universally characterize fractional excitations in a given foliated fracton phase. A quotient superselection sector (QSS) is defined as an equivalence class of ordinary superselection sectors modulo the planon superselection sectors that come from the resource layers used in the RG procedure. In other words, two point-like fractional excitations belong to the same QSS if they are related to each other through local operations and the addition or removal of planon excitations that are unitarily equivalent to anyons in the resource layers. In the Majorana checkerboard model, all planons are transformed into toric code anyons under the inverse RG transformation of Sec. \ref{sec:rg}.  To see this, note that the planon string operators are mapped into toric code string operators under the inverse RG transformation $S^\dagger$.

To describe the QSS of the Majorana checkerboard model, it is helpful to further partition the $A$ checkerboard sublattice into 4 sublattices labelled $R$, $G$, $B$, and $Y$, as in Fig. \ref{fig:unitcelllabelled}. Excited states may be labelled according to which Hamiltonian stabilizers they violate (e.g. the error syndrome). Planon excitations violate two stabilizers corresponding to adjacent sites of either the $R$, $G$, $B$, or $Y$ sublattice. For instance, the planon depicted in Fig. \ref{fig:majoranaexcitation} violates two adjacent $B$ sublattice Hamiltonian terms. Thus, the addition of planons on a given sublattice acts as a pair creation/annihilation, or hopping, operator for excitations of the stabilizers on that sublattice. As a result, we find that the QSS are characterized by the parity of the error syndrome on each sublattice, and can be labelled accordingly. For instance, the lineon depicted in Fig. \ref{fig:majoranaexcitation} belongs to the $RB$ QSS because the state violates one $R$ stabilizer and one $B$ stabilizer. However, since a local fermionic excitation corresponds to a violation of one stabilizer of each of the $R$, $G$, $B$, and $Y$ sublattices, the $RGBY$ QSS is in fact identified with the vacuum sector (and $RB$ is identified with $GY$, and so forth). Therefore, a complete list of the 8 QSS is given in the first column of Table I.

\begin{figure}[htbp]
    \centering
    \includegraphics[width=0.4\textwidth]{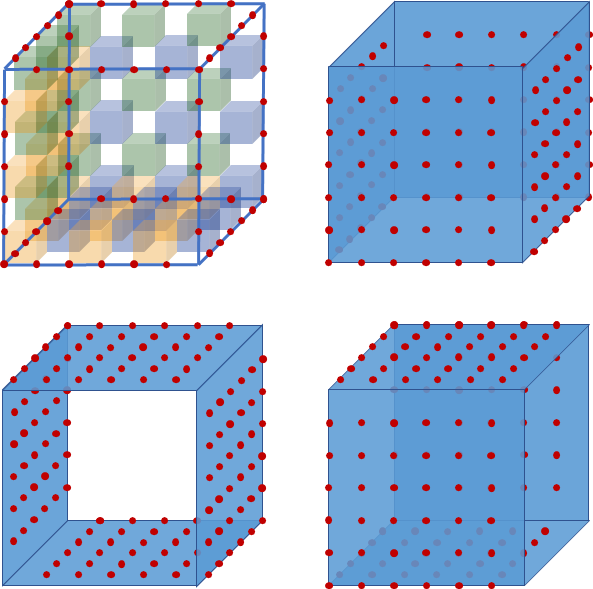}
    \caption{Interferometric operators in the Majorana checkerboard, which correspond to products of Majoranas over the red sites. $RGBY$ wireframe operator (top left), and $BY$, $GB$, and $GY$ cylindrical membrane operators (top right, bottom left, bottom right).}
    \label{fig:interferometric}
\end{figure}

\begin{table}[hbtp]
\label{tab:correspondence}
\begin{tabular}{|c|c||c|c|}
    \hline
    Majorana QSS & X-cube QSS & Majorana IOs & X-cube IOs\\ 
    \hline
    $1$ & $1$ & $1$ & $1$\\
    $R$ & $f$ & $RGBY$ & $F$\\ 
    $BY$ & $\ell_x$ & $BY$ & $X$\\ 
    $GB$ & $\ell_y$ & $GB$ & $Y$\\ 
    $GY$ & $\ell_z$ & $GY$ & $Z$\\ 
    $RBY$ & $\ell_x \times f$ & $RG$ & $XF$\\ 
    $RGB$ & $\ell_y \times f$ & $RY$ & $YF$\\ 
    $RGY$ & $\ell_z \times f$ & $RB$ & $ZF$\\
    \hline
\end{tabular}
\caption{Correspondences between the quotient superselection sectors (QSS) and interferometric operators (IOs) of the Majorana checkerboard and X-cube models.}
\end{table}

In fact, in terms of the mobility of the excitations and their fusion rules, there is an exact correspondence between the QSS of the Majorana checkerboard model and those of the X-cube foliated fracton phase, given in the table. In particular, the three lineon sectors of the X-cube model correspond to the $BY$, $GB$, and $GY$ lineon sectors of the Majorana checkerboard model, which likewise obey a triple fusion rule. On the other hand, the $R$, $G$ ($RBY$), $B$ ($RGY$), and $Y$ ($RGB$) fracton sectors correspond to the fractonic sectors $f$, $f\times \ell_x$, $f\times \ell_y$, and $f\times \ell_z$ of the X-cube model. Of course, there is an ambiguity as to which of the Majorana checkerboard fracton sectors is chosen to correspond to the $f$ sector. In our case we have chosen the $R$ sector. As we will see in the following sections, this correspondence must exist due to the local unitary equivalence of the model with a fermionic version of the semionic X-cube model, which is known to lie in the X-cube foliated fracton phase.

Ref. \onlinecite{FractonStatistics} also introduced the notion of \textit{interferometric operators}, which are classes of unitary operators that detect the QSS content of a given region but are insensitive to the planon content of the region. The equivalence of the foliated fracton order in the Majorana checkerboard model with that of the X-cube model manifests not only as a correspondence between QSS, but furthermore as a correspondence between the interferometric operators of the two models. As discussed in Ref. \onlinecite{FractonStatistics}, there are 8 classes of interferometric operators for the X-cube model, which include a wireframe operator $F$ and three cylinder membrane operators $X$, $Y$, and $Z$ (whose axes lie along the $x$, $y$, and $z$ directions), and the composites $XF$, $YF$, and $ZF$. Each of these classes corresponds to a class of operators in the Majorana checkerboard model whose regions of support have the identical geometry (wireframe or cylinder with axis along the $x$, $y$, or $z$ direction) and whose interferometric statistics agree exactly with the corresponding statistics of the X-cube model.

These interferometric operators can be written as products of Hamiltonian terms within a large cubic region. In particular, we will denote by $RGBY$ the product of all $R$, $G$, $B$, and $Y$ cube terms within the large cubic region, by $BY$ the product of all $B$ and $Y$ cube terms, and so on and so forth. In this notation, the wireframe operator corresponds to $RGBY$ whereas the 3 cylindrical membrane operators correspond to $BY$, $GB$, and $GY$ respectively. These operators are illustrated in Fig. \ref{fig:interferometric}, and the full correspondence is given in the table above. As an example, the $X$ membrane operator yields a $\pi$ phase when it acts on a state with quotient charge $\ell_y$, $\ell_z$, $f\ell_y$, or $f\ell_z$. Correspondingly, the $BY$ membrane operator has a $\pi$ statistic with the $GB$, $GY$, $RGB$, and $RGY$ quotient sectors.

\section{Mapping the Majorana Checkerboard Model to a Spin Model} \label{sec:mapping1}

\subsection{Mapping to a spin model}

In this section, we describe a local unitary transformation from the Majorana checkerboard model to a bosonic stabilizer code augmented with decoupled fermionic degrees of freedom. A mapping of the same spirit between the Majorana color code on the square-octagon lattice \cite{PhysRevX.5.041038,MFC} and the Wen plaquette model plus decoupled fermions on a square lattice \cite{PhysRevLett.90.016803} is briefly discussed in Appendix \ref{sec:2dmajoranamapping}.

\begin{figure}[htbp]
    \centering
    \includegraphics[width=0.4\textwidth]{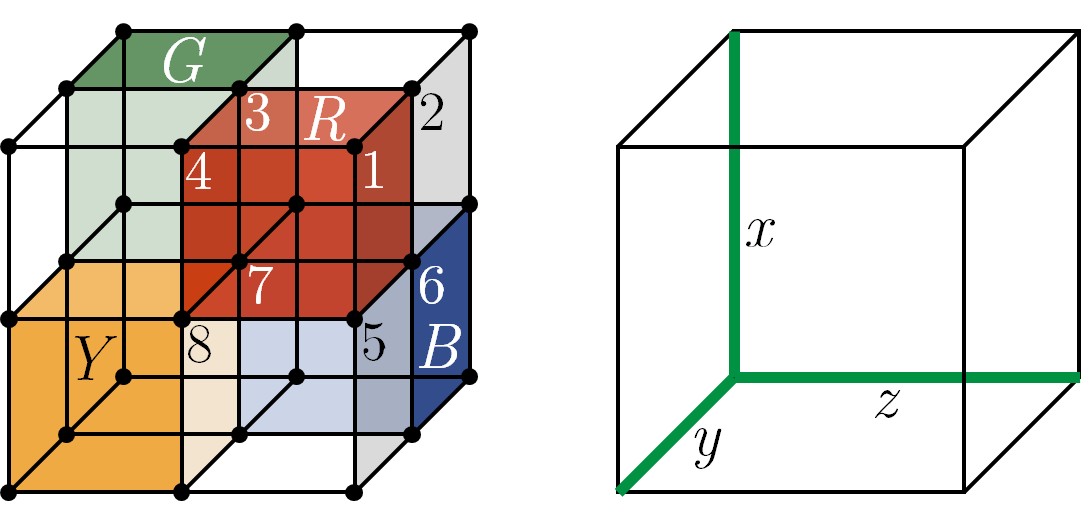}
    \caption{(Left) Unit cell of the Majorana checkerboard model with the Majorana degrees of freedom labelled from $1$ to $8$. The unit cell contains one cube of each of the $R$, $G$, $B$, and $Y$ sublattices. (Right) Unit cell of the spin model containing a qubit degree of freedom on the green edges labelled $x$, $y$ and $z$.}
    \label{fig:unitcelllabelled}
\end{figure}

For our purposes we consider a unit cell of the Majorana checkerboard model as a $2\times2\times2$ cell of the underlying cubic lattice, which contains one cube of each of the $R$, $G$, $B$, and $Y$ sublattices and 8 Majorana fermion degrees of freedom, labelled as shown in Fig. \ref{fig:unitcelllabelled}(a). The spin model we consider has one qubit degree of freedom on each edge of a cubic lattice, and thus has 3 qubits per unit cell, which are labelled according to the direction of the edge as in Fig. \ref{fig:unitcelllabelled}(b). This bosonic Hilbert space augmented with 2 Majorana fermions per unit cell, labelled $\gamma_A$ and $\gamma_B$, is identical to the Hilbert space of the Majorana checkerboard model (each being 16-dimensional in a unit cell). We describe a parity-preserving local unitary transformation $U^\dagger$ from the composite spin and Majorana Hilbert space to the pure Majorana Hilbert space via its action on the generators of the operator algebra. In particular, within each unit cell, $U^\dagger$ maps
\begin{gather}
    X^x \to \gamma_1\gamma_5, \quad Z^x \to \gamma_5\gamma_6\gamma_7\gamma_8\\ X^y \to \gamma_3\gamma_4, \quad Z^y \to \gamma_2\gamma_3\gamma_6\gamma_7\\ X^z \to \gamma_6\gamma_7, \quad Z^z \to \gamma_3\gamma_4\gamma_7\gamma_8\\
    \gamma_A\to \gamma_2\gamma_3\gamma_4, \quad \gamma_B\to \gamma_1\gamma_5\gamma_6\gamma_7\gamma_8.
\end{gather}
Note that the commutation relations of the algebra are preserved as well as the global fermionic parity.

\begin{figure*}[hbtp]
    \centering
    \includegraphics[width=.95\textwidth]{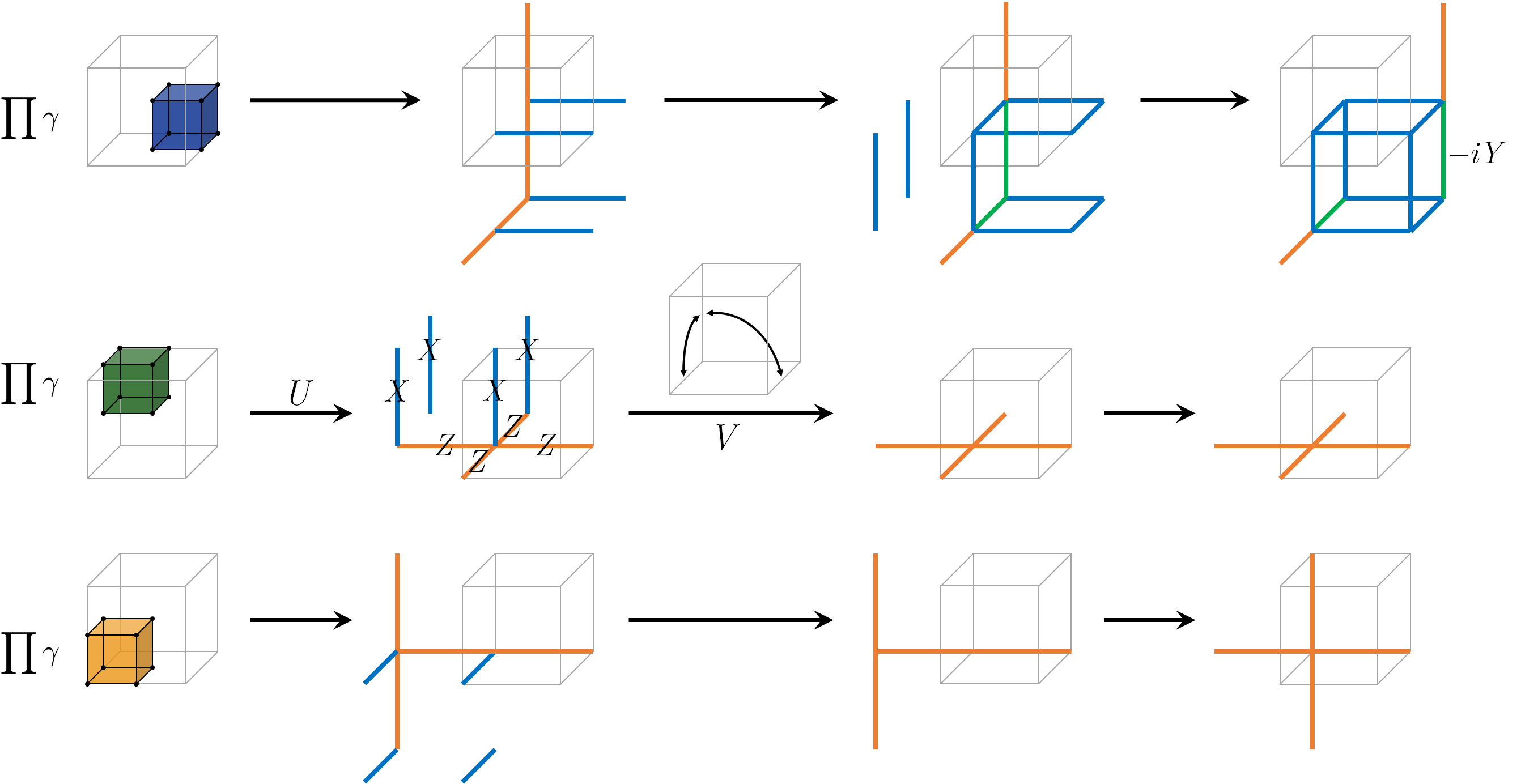}
    \caption{Mapping from the green, blue, and yellow sublattice cube terms of the Majorana checkerboard Hamiltonian to the stabilizer terms of the new spin Hamiltonian $H^0_\mathrm{spin}$. The spin stabilizers are tensor products of Pauli operators acting on the qubits on the colored edges: blue for Pauli $X$, green for $-iY=XZ$, and orange for $Z$. The first step is the unitary $U$, whereas the second step is the unitary $V$. A unit cell of $V$ is depicted in the inset, where an arrow between two qubits represents the gate $H(CZ)H$. The final step of the transformation is simply a redefinition of the unit cell.}
    \label{fig:majoranaspinmapping}
\end{figure*}

The $G$, $B$, and $Y$ sublattice stabilizer terms of the Majorana checkerboard model are transformed under $U$ to the bosonic stabilizers shown in Fig. \ref{fig:majoranaspinmapping}, whereas the $R$ sublattice terms map to the local parity check $\gamma_A\gamma_B$. Therefore $U$ decouples the system into a bosonic stabilizer code and a trivial Majorana stabilizer code.

The bosonic code can be further massaged into a more amenable form. In particular, consider the local unitary operator
\begin{equation}
    V=H\left(\prod_{i} CZ^{i,z}_{i,y} CZ^{i,z}_{i+\hat{x},y}\right)H,
\end{equation}
where the index $i$ runs over all unit cells of the underlying cubic lattice, the operator $CZ^{i,\mu}_{j,\nu}$ is the controlled-$Z$ operator acting on the $\mu$-oriented edge of unit cell $i$ and the $\nu$-oriented edge of unit cell $j$, and $H$ is a global Hadamard rotation. The unitary $V$ is depicted graphically in Fig. \ref{fig:majoranaspinmapping}.
Under conjugation by $H(CZ)H$, the two-qubit Pauli operators transform as follows:
\begin{equation}
\begin{split}
    XI \to XI,\quad IX \to IX,\\
    ZI \leftrightarrow ZX,\quad IZ \leftrightarrow XZ.
\end{split}
\end{equation}
Hence, the stabilizers of the qubit stabilizer code are transformed under $V$ as shown in Fig. \ref{fig:majoranaspinmapping}. Finally, it is convenient to redefine the unit cell by shifting the vertical edges by one unit to the right, thus yielding the stabilizer terms on the far right side of Fig. \ref{fig:majoranaspinmapping}. Let us denote the Hamiltonian corresponding to these stabilizers as $H^0_\mathrm{spin}$. In summary, we find that
\begin{equation}
    (UV) H (UV)^\dagger
    \cong H^0_\mathrm{spin}+H_f,
\end{equation}
where $H$ is the Majorana checkerboard Hamiltonian and $H_f=-i\sum\gamma_A\gamma_B$ stabilizes the ancillary Majorana degrees of freedom. Here the relation $\cong$ denotes that the two Hamiltonians have identical ground spaces.

\subsection{Analysis of the spin model}

\begin{figure}[htbp]
    \centering
    \includegraphics[width=0.4\textwidth]{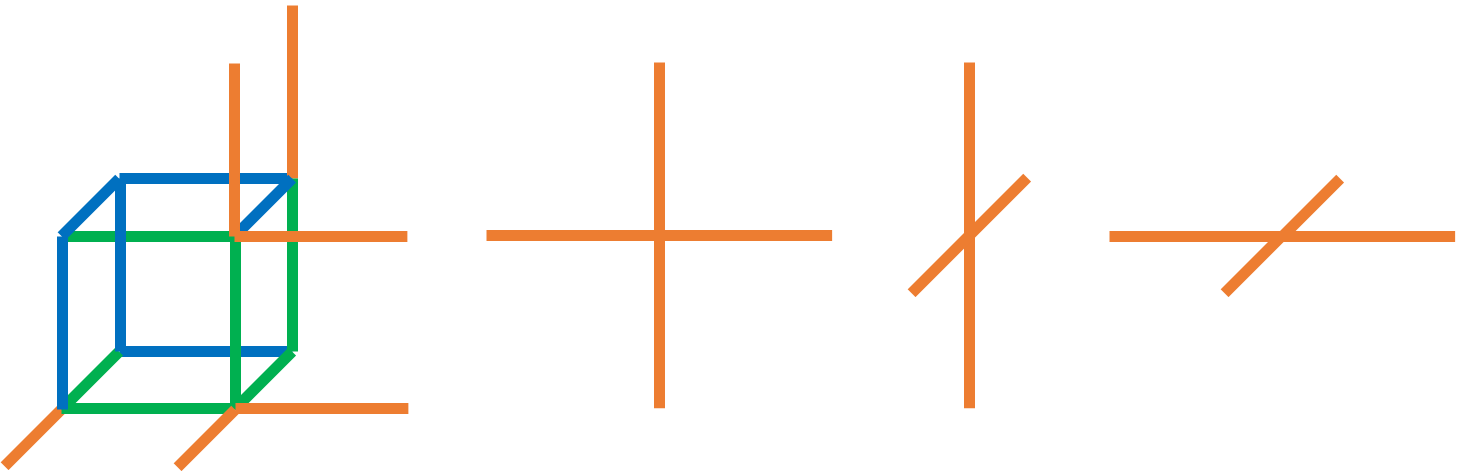}
    \caption{Cube ($B_c^\mathrm{spin}$, left) and vertex ($A^\mu_v$, right) terms of the stabilizer code Hamiltonian $H_\mathrm{spin}$. The stabilizers are tensor products of Pauli operators acting on the qubits on the colored edges: blue for Pauli $X$, green for $-iY=XZ$, and orange for $Z$.}
    \label{fig:spinH}
\end{figure}

It is instructive to consider a Hamiltonian $H_\mathrm{spin}$ which is equivalent as a stabilizer code to $H^0_\mathrm{spin}$, but whose form is analogous to that of the X-cube model.\cite{Sagar16} This representation will highlight the differences between this spin model and the X-cube model; as we will see in the next section, the model is in fact a stabilizer code realization of the \textit{semionic} X-cube model.\cite{MaLayers} In particular, we define
\begin{equation}
    H_\mathrm{spin}=-\sum_v\left(A_v^x+A_v^y+A_v^z\right)-\sum_cB_c^\mathrm{spin}
\end{equation}
where $v$ runs over all vertices and $c$ over all elementary cubes. Here $A_v^\mu$ are vertex terms and $B_c^\mathrm{spin}$ is a cube term, as depicted in Fig. \ref{fig:spinH}. Note that $B_c^\mathrm{spin}$ can be decomposed as a product of Pauli $Z$ operators followed by the product of Pauli $X$ operators over the 12 edges of the cube $c$. The vertex terms are identical to those of the X-cube model, whereas the cube term differs inasmuch as it contain factors of $Z$ operators in addition to the product of $X$ operators. Note that $H_\textrm{spin}$ indeed generates the same stabilizer group as $H_\textrm{spin}^0$: the additional vertex term is generated by the other two vertex terms and hence redundant, whereas $B_c^\textrm{spin}$ is generated by the stabilizer in the top right corner of Fig. \ref{fig:majoranaspinmapping} along with two nearby vertex terms. The fractional excitations of the model can be organized into fracton and lineon sectors, which respectively correspond to violations of the cube and vertex terms.

The fracton sector of $H_\mathrm{spin}$ is identical to the fracton sector of the X-cube model. In particular, fractons are created at the corners of rectangular membrane operators, which are products of Pauli $Z$ operators and hence commute with all vertex terms but anti-commute with the cube stabilizers at the corners of the membrane. Moreover, fracton dipoles, which are composites of adjacent fracton excitations, are planons, as in the X-cube model.

\begin{figure}[htbp]
    \centering
    \includegraphics[width=0.3\textwidth]{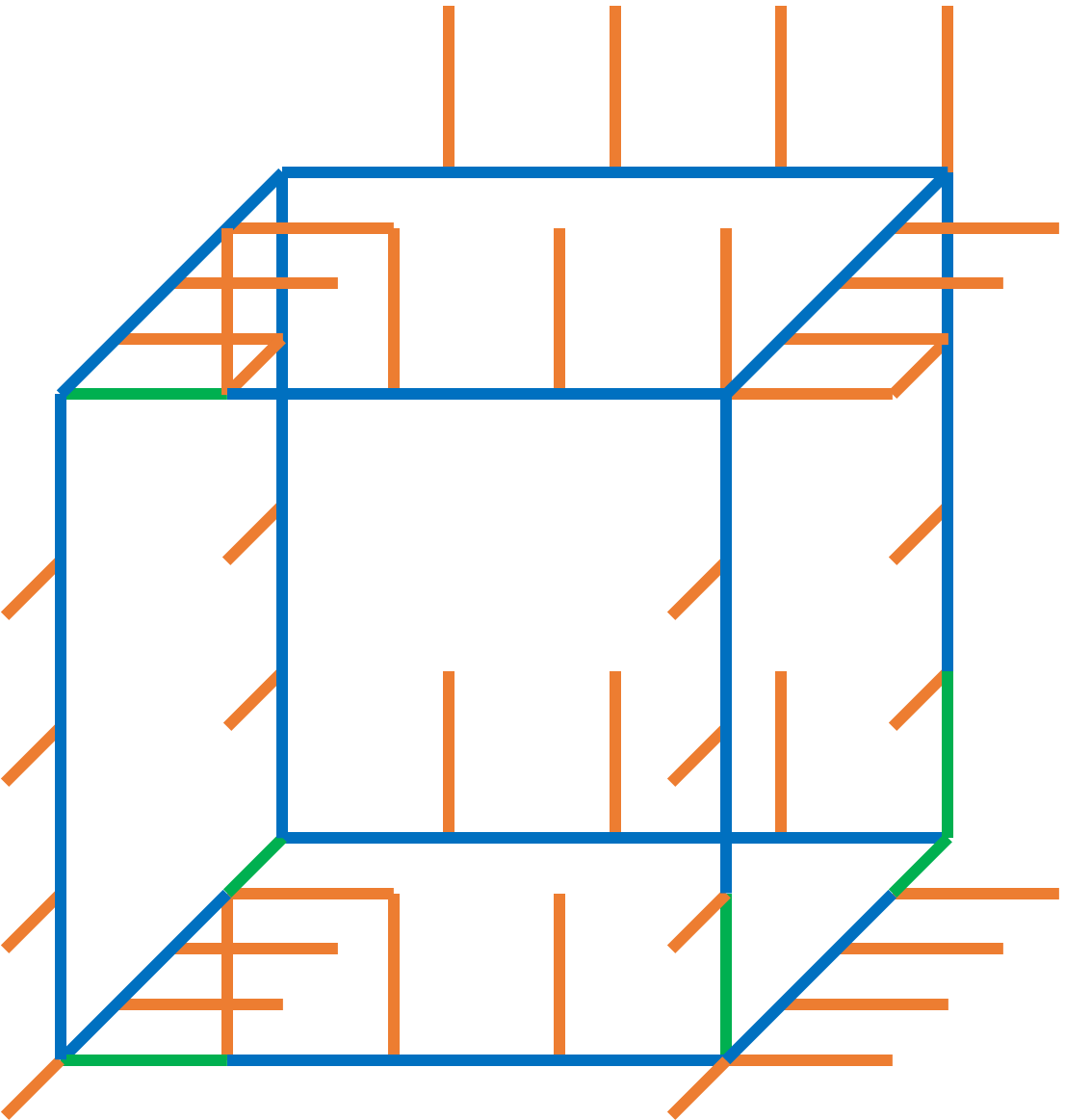}
    \caption{Wireframe operator of the spin model $H_\mathrm{spin}$, which is equal to the product of cube terms $B_c^\mathrm{spin}$ within the large cubic region.}
    \label{fig:wireframe}
\end{figure}

Conversely, the lineon sector of the model is subtly different from that of the X-cube model. As in the X-cube model, the product of all cube terms $B_c^\mathrm{spin}$ within a large cubic region yields a large operator with support near the wireframe of the large cubic region, as depicted in Fig. \ref{fig:wireframe}. (It is for this reason that we have chosen the particular form of $B_c^\mathrm{spin}$). In fact, this wireframe operator corresponds to a physical process in which lineons travel along all of the edges of the cube, fusing and splitting at the corners according to triple fusion rules in which a lineon in each of the $x$, $y$, and $z$ directions come together and annihilate into the vacuum. Thus, the rigid string operators which transport lineons in this model have the same form as the edges of the wireframe operator.

From this observation, it becomes clear by inspecting the wireframe operator in Fig. \ref{fig:wireframe} that pairs of perpendicularly-moving lineons which are involved in a triple fusion rule have a mutual `semionic braiding' statistic, in the sense that the rigid string operators which create these lineons \textit{anti-commute} with each other. This property lies in stark contrast to the X-cube model where lineons satisfying a triple fusion rule always have trivial mutual `braiding'. In fact, this characteristic is the only essential difference between the X-cube model and the spin model here.

The structure of non-local excitations in $H_\mathrm{spin}$ is highly reminiscent of the discussion of quasiparticles in the \textit{semionic X-cube model} of Ref. \onlinecite{MaLayers}. Indeed, it was shown that that model differs fundamentally from the X-cube model only insofar as lineons satisfying a triple fusion rule have mutually anti-commuting, as opposed to commuting, string operators. Therefore, we see that in fact the semionic X-cube model and our spin Hamiltonian have isomorphic structures of non-local excitations in terms of fusion and braiding. It is thus natural to expect that they are in fact equivalent models under local unitary transformation. We will see in the next section an explicit description of such a transformation.

\section{Mapping the Spin Model to the Semionic X-Cube Model} \label{sec:mapping2}

In this section, we describe a local unitary transformation between the ground spaces of the semionic X-cube model and the stabilizer code spin model $H_\mathrm{spin}$ obtained in the previous section.

\subsection{Semionic X-cube model}

\begin{figure}[htbp]
    \centering
    \subfloat[]{
    \includegraphics[width=0.26\textwidth]{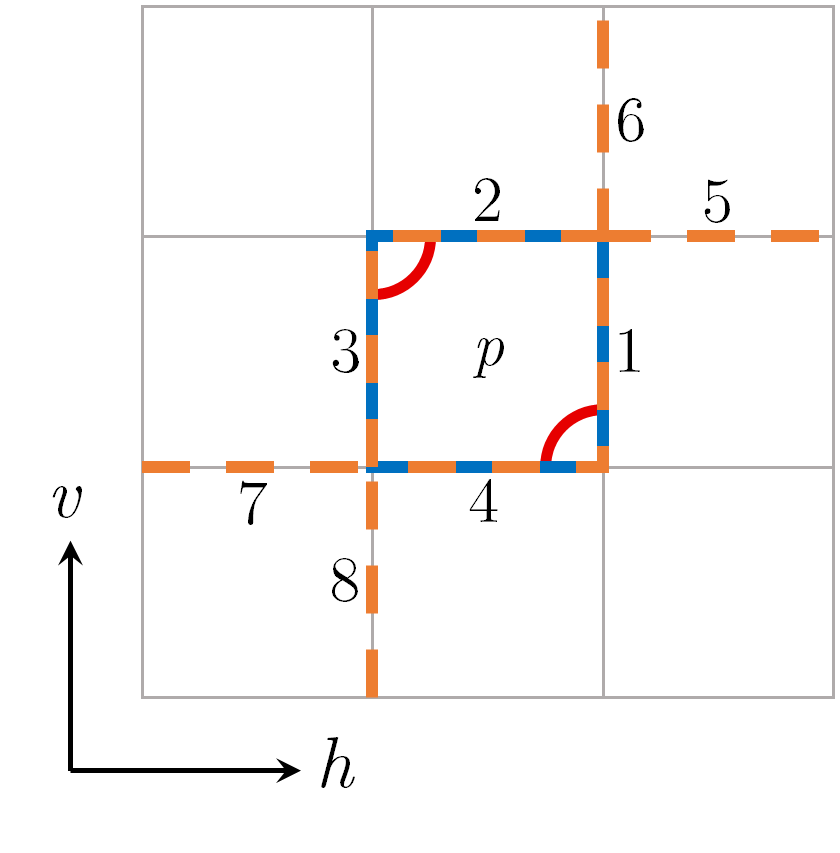}}
    \qquad
    \subfloat[]{
    \includegraphics[width=0.15\textwidth]{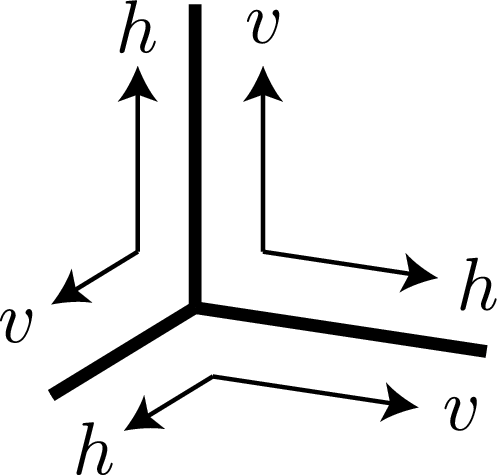}}
    \caption{(a) The component $B_p$ of the double semion model plaquette term $\tilde{B}_p$. Here, dashed orange edges represent the phase gate $S = i^\frac{1-Z}{2}$, blue-orange dashed edges represent the operator $XS$ and the red arcs represents the controlled-$Z$ gate between the two linked edges. The action of the $CZ$ gates precede the action of the $XS$ operators. (b) The orientations of the double semion layers in the three stacks prior to coupling.}
    \label{fig:couple}
\end{figure}

The semionic X-cube model, as first discussed in Ref. \onlinecite{MaLayers} is obtained by coupling together three mutually perpendicular interpenetrating stacks of 2D double semion models\cite{Stringnet} on the square-octagon lattice. For our purposes, it is more convenient to work with a microscopic realization of the double semion model whose degrees of freedom are qubits on a square lattice (see Appendix \ref{sec:ds}). The Hamiltonian takes the form
\begin{equation}
    H_\mathrm{DS}=-\sum_v A_v-\sum_p \tilde{B}_p
\end{equation}
where $v$ runs over all vertices of the square lattice and $p$ runs over all plaquettes. The vertex term $A_v$ is defined as the product of Pauli $Z$ operators over the edges adjacent to $v$, whereas the plaquette term $B_p$ is defined as follows:
\begin{equation}
    \tilde{B}_p=B_p\prod_{v\in p} \frac{1+A_v}{2},
\end{equation}
where $v$ runs over the vertices surrounding plaquette $p$ and $B_p$ is a unitary operator which is depicted graphically in Fig. \ref{fig:couple}(a). Explicitly,
\begin{equation}
    B_p=X_1X_2X_3X_4S_1S_2S_3S_4S_5S_6S_7S_8CZ_{14}CZ_{23}
\end{equation}
where the qubits are numbered as in Fig. \ref{fig:couple}(a). Here $CZ_{ij}$ denotes the controlled-$Z$ gate between qubits $i$ and $j$ and $S=i^\frac{1-Z}{2}=\mathrm{diag}(1,i)$). 

\begin{figure*}[htbp]
    \centering
    \includegraphics[width=\textwidth]{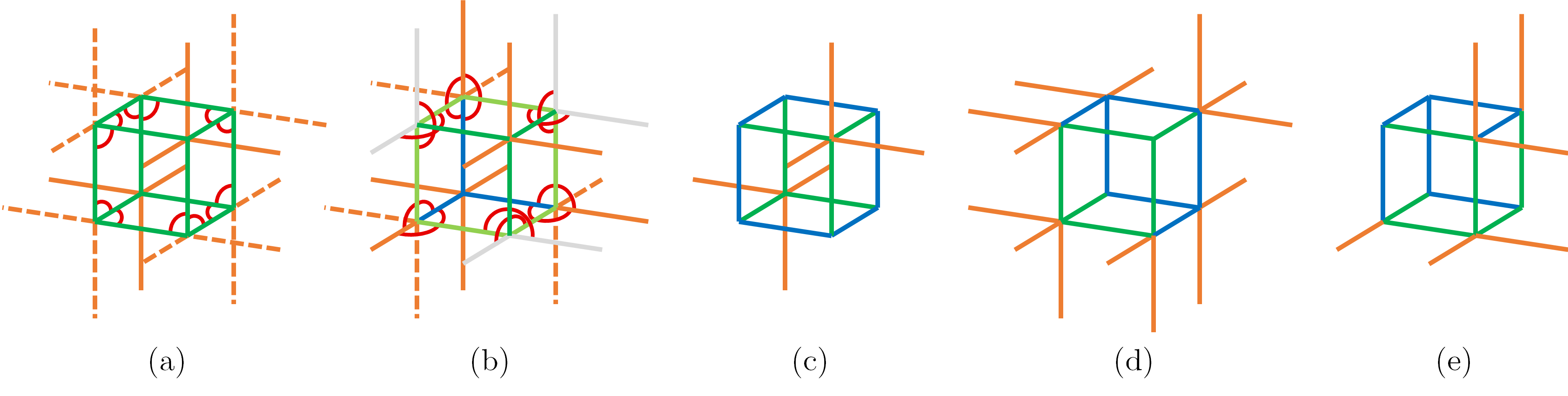}
    \caption{Graphical depictions of the operators (a) $B_c^a$, (b) $B_c^b$, (c) $B_c^c$, (d) $B_c^d$, and (e) $B_c^e$. Operators $B_c^c$, $B_c^d$, and $B_c^e$ are simply tensor products of Pauli operators acting on the qubits on the colored edges: blue for Pauli $X$, green for $-iY=XZ$, and orange for $Z$. Conversely, $B_c^a$ and $B_c^b$ are each composed of two pieces: first, the tensor product of the controlled-$Z$ two-qubit gates depicted as red arcs linking the two qubits. Second, the tensor product of single-qubit gates illustrated: blue, orange, and green for the Pauli operators, dashed orange for the phase gate $S$, and light green edges for the operator $XSZ$. The gray edges are simply placeholders.}
    \label{fig:semionxchamiltonian}
\end{figure*}

To obtain the semionic X-cube model, we consider three stacks of double semion layers in the $x$, $y$, and $z$ directions, whose edges coincide with the edges of a cubic lattice. The layers in the stack are oriented as illustrated in Fig. \ref{fig:couple}(b). Each edge thus lies at the intersection of two double semion layers, and contains two qubit degrees of freedom. The two qubits on each edge are subsequently subjected to a $ZZ$ coupling. To be precise, we consider the following Hamiltonian:
\begin{equation}
    H = \sum_L H^L_{DS} - J\sum_{e}Z^{\mu_1}_eZ^{\mu_2}_e,
\end{equation}
where $L$ indexes the layers of all three stacks, $e$ runs over all edges of the cubic lattice, $H^L_{DS}$ is the double semion Hamiltonian in layer $L$, and $Z_{\mu_1}^1$ and $Z_{\mu_2}^1$ are Pauli operators acting on the two qubits on edge $e$. In the strong coupling limit $J \to \infty$, the two qubits on each edge effectively combine into one degree of freedom. The effective Hamiltonian to leading order in $1/J$ is given by
\begin{equation}
    H_\mathrm{sem}=-\sum_v\left(A_v^x+A_v^y+A_v^z\right)-\sum_c\tilde{B}_c^\mathrm{sem},
\end{equation}
where the vertex terms $A_v^\mu$ are the same as those of the X-cube model and $H_\mathrm{spin}$. In fact, note that this Hamiltonian is identical to $H_\mathrm{spin}$ apart from the cube term $\tilde{B}_c^\mathrm{sem}$. The cube term $\tilde{B}_c^\mathrm{sem}$ can be written as
\begin{equation}
    \tilde{B}_c^\mathrm{sem}=B_c^a\prod_{v\in c}\prod_{\mu=x,y,z} \frac{1+A^\mu_v}{2}.
\end{equation}
Here the factors on the right-hand side project into the subspace satisfying the vertex constraints at the corners of the cube $c$. The unitary operator ${B}_c^a$ is depicted graphically in Fig. \ref{fig:semionxchamiltonian}(a). It can be decomposed as a unitary operator diagonal in the Pauli $Z$ basis followed by a product of the Pauli $X$ operators around the 12 edges of the cube $c$.

\subsection{Mapping to $H_\mathrm{spin}$}

First, let us define a modified spin Hamiltonian $\tilde{H}_\mathrm{spin}$ which is identical to $H_\mathrm{spin}$ except for the replacement $B_c^\mathrm{spin}\to \tilde{B}_c^\mathrm{spin}$ where
\begin{equation}
    \tilde{B}_c^\mathrm{spin}=B_c^\mathrm{spin}\prod_{v\in c}\prod_{\mu=x,y,z}\frac{1+A^\mu_v}{2}.
\end{equation}
Here $v$ runs over the corners of the cube $c$. Since the additional factors on the right-hand side simply project into the subspace satisfying all of the vertex constraints around $c$, it is clear that $\tilde{H}_\mathrm{spin}$ has the same ground space as the stabilizer code $H_\mathrm{spin}$. We will now describe a local unitary operator $W$ such that $W^\dagger H_\mathrm{sem}W=\tilde{H}_\mathrm{spin}$, demonstrating that $H_\mathrm{spin}$ is in fact a stabilizer code realization of the semionic X-cube model.

\begin{figure}[htbp]
    \centering
    \includegraphics[width=.35\textwidth]{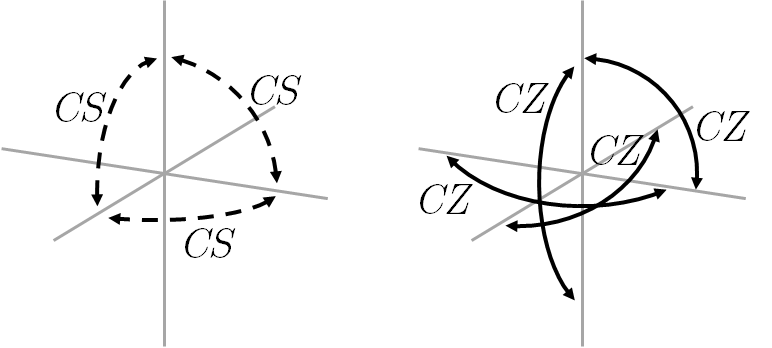}
    \caption{Illustration of a unit cell of the unitary operators $W_1$ (left) and $W_2$ (right). Here the dashed arrows represent controlled-phase gates between the two endpoints, whereas the solid arrows represent controlled-$Z$ gates.}
    \label{fig:W1W2}
\end{figure}

\begin{figure}[htbp]
    \centering
    \includegraphics[width=.3\textwidth]{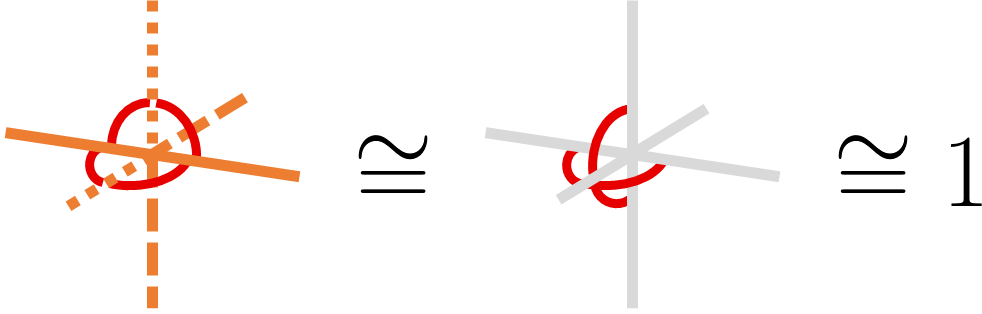}
    \caption{Operator relations that hold within the subspace satisfying the vertex constraints. These relations can be used to equate $B_c^b$ and $B_c^c$ within this subspace. Here, the red arcs represent controlled-$Z$ gates, solid orange represents $Z$, dashed orange represents $S$, and dotted orange represents $S^\dagger=SZ$.}
    \label{fig:vertexrelations}
\end{figure}

The operator $W$ can be decomposed as $W=W_2W_1$ where $W_1$ and $W_2$ are both unitary. Consider as a unit cell the three edges depicted on the right-hand side of Fig. \ref{fig:unitcelllabelled}(b). The first factor $W_1$ is defined as
\begin{equation}
    W_1=\prod_i \left(CS^{i,x}_{i,y}\times CS^{i,y}_{i,z}\times CS^{i,z}_{i,x}\right)
\end{equation}
where $CS^{i,\mu}_{j,\nu}$ is a controlled-phase gate between the $\mu$-oriented edge in unit cell $i$ and the $\nu$-oriented edge in unit cell $j$, and the index $i$ runs over all unit cells (see Fig. \ref{fig:W1W2}). In matrix form, $CS = \mathrm{diag}(1,1,1,i)$. The action of $CS$ by conjugation is given by
\begin{align}
    X_1 &\to X_1 S_2 CZ_{12}\\
    X_2 &\to X_2S_1CZ_{12}
\end{align}
where $CZ_{12}$ is the controlled-$Z$ gate acting on qubits $1$ and $2$, and $S_1$ ($X_1$) and $S_2$ ($X_2$) are the $S$ ($X$) operators acting on qubits 1 and 2 respectively. It hence follows that $W_1^\dagger B_c^a W_1=B_c^b$, where $B_c^b$ is the operator depicted in Fig. \ref{fig:semionxchamiltonian}(b). Furthermore, since $B_c^b$ is equivalent to $B_c^c$ within the subspace satisfying the vertex constraints around $c$ (see Fig. \ref{fig:vertexrelations}), it follows that
\begin{equation}
    W_1^\dagger\tilde{B}_c^\mathrm{sem}W_1=B_c^c\prod_{v\in c}\prod_{\mu=x,y,z} \frac{1+A^\mu_v}{2}.
\end{equation}
Here $B_c^c$ is the operator depicted in Fig. \ref{fig:semionxchamiltonian}(c).

The second factor $W_2$ is defined as (see Fig. \ref{fig:W1W2})
\begin{equation}
    W_2=\prod_i \left(CZ^{i,x}_{i+\hat{x},x}\times CZ^{i,y}_{i+\hat{y},y}\times CZ^{i,z}_{i+\hat{z},z} \times CZ^{i,x}_{i,z}\right)
\end{equation}
where $CZ^{i,\mu}_{j,\nu}$ is a controlled-$Z$ gate between the $\mu$-oriented edge in unit cell $i$ and the $\nu$-oriented edge in unit cell $j$, and the index $i$ runs over all unit cells. Since $CZ$ acts by conjugation as
\begin{equation}
\begin{split}
    XI \to XI,\quad IX \to IX,\\
    ZI \leftrightarrow ZX,\quad IZ \leftrightarrow XZ,
\end{split}
\end{equation}
it follows that $W_2^\dagger B_c^c W_2=B_c^d$, where $B_c^d$ is depicted graphically in Fig. \ref{fig:semionxchamiltonian}(d). Finally, this yields the result
\begin{equation}
    W^\dagger\tilde{B}_c^\mathrm{sem}W=B_c^e\prod_{v\in c}\prod_{\mu=x,y,z} \frac{1+A^\mu_v}{2}
\end{equation}
due to the equivalence of $B_c^d$ and $B_c^e$ within the projected subspace. The unitary $B_c^e$ is depicted in Fig. \ref{fig:semionxchamiltonian}(e). Since $B_c^e$=$B_c^\mathrm{spin}$, it thus follows that $W^\dagger \tilde{B}_c^\mathrm{sem} W=\tilde{B}_c^\mathrm{spin}$. Since $W$ is diagonal in the $Z$ basis, it leaves the vertex terms unaffected, and hence altogether $W^\dagger H_\mathrm{sem}W=\tilde{H}_\mathrm{spin}$.

We have therefore verified the intuitive correspondence between the Majorana checkerboard model and the semionic X-cube model (plus decoupled fermionic modes) by explicitly describing a local unitary transformation between the two models. As an intermediate step we have demonstrated how to decouple the fermionic degrees of freedom of the Majorana checkerboard model from a hidden bosonic stabilizer code representation of the semionic X-cube model. Indeed, in light of the exact correspondence between the structure of non-local excitations of $H_\mathrm{spin}$ and $H_\mathrm{sem}$, the existence of such a local unitary equivalence is to be expected.

In a previous work, it was demonstrated that the semionic X-cube model lies in the same foliated fracton phase as the X-cube model.\cite{FractonStatistics} Indeed, the anti-commutation of string operators which satisfy a triple fusion rule in the semionic X-cube model can be completely cancelled by the addition of three mutually perpendicular stacks of 2D double semion layers. Consequently, the result of the current work implies that the Majorana checkerboard model too lies in the X-cube foliated fracton phase.

\section{Discussion}
\label{sec:discussion}

To summarize, we have shown in this paper that the Majorana checkerboard model, first introduced in Ref. \onlinecite{VijayFracton}, has foliated fracton order as defined in Ref. \onlinecite{3manifolds,FractonEntanglement}. That is, 2D topological states are extracted from the bulk when renormalization group transformations are applied to the ground state wavefunction to reduce the total system size. Moreover, we show through explicit mapping that the Majorana checkerboard model has the same foliated fracton order as the X-cube model. This equivalence may not be straightforward to see given the many differences between the two models: The Majorana checkerboard model is fermionic while the X-cube model is bosonic; moreover, the Majorana checkerboard model has a `dimensional hierarchy' of quasiparticle fusion while this does not seem to be the case in the X-cube model. By calculating the universal properties of foliated fracton phases as discussed in Ref. \onlinecite{FractonEntanglement,FractonStatistics,Gauging,Checkerboard}, we see that the two models could actually be in the same foliated fracton phase, and the explicit mapping discussed in section \ref{sec:mapping1} and section \ref{sec:mapping2} further confirms this result.

So far we have found, using the same procedure as in this paper, phase relations between several type I fracton models including the X-cube model, the checkerboard model (as two copies of X-cube)\cite{Checkerboard}, the semionic X-cube model\cite{FractonStatistics} and the Majorana checkerboard model. These models all belong to the same foliated fracton phase. On the other hand, other types of foliated fracton phase can also exist. We have found that some Type-I fracton models have foliated fracton order distinct from that of the X-cube model. These results will be presented in a separate work.\footnote{W. Shirley, K. Slagle, and X. Chen, forthcoming.}

\begin{acknowledgments}
We are grateful to Kevin Slagle for helpful discussions. T.W. is supported by Caltech's Summer Undergraduate Research Fellowships and the Institute for Quantum Information and Matter at Caltech. W.S. and X.C. are supported by the National Science Foundation
under award number DMR-1654340 and the Institute for Quantum Information and Matter at Caltech. X.C. is also supported by the Alfred P. Sloan research fellowship and the Walter
Burke Institute for Theoretical Physics at Caltech.
\end{acknowledgments}

\appendix

\section{Entanglement entropy in Majorana codes} \label{sec:majoranaentanglement}

In this appendix, we show that the entanglement entropy of a subregion of a Majorana code is equal to half that of the corresponding `doubled' CSS code. This self-dual CSS code is constructed by replacing each Majorana fermion with a qubit, and each Majorana stabilizer by one $X$ type qubit stabilizer and one $Z$ type qubit stabilizer.\cite{MFC} For instance, the spin checkerboard model arises as the `double' of the Majorana checkerboard model. The method of calculation straightforwardly generalizes that of qubit stabilizer codes.\cite{entanglementStabilizer}

Consider a Majorana code with stabilizer group $S$ generated by $n$ independent commuting Majorana stabilizer operators $g_1,\ldots,g_n$ on a Hilbert space of $2n$ Majorana modes. The stabilizers are of the form $g_i = \prod_{j\in S_i} i^{1/2} \gamma_j$, where $S_i$ labels the support of $g_i$. To calculate the ground state entanglement entropy of a subregion $A$, the ground state density matrix $\rho=\ket{\psi}\bra{\psi}$ may be written as
\begin{equation}
    \rho = \frac{1}{2^{2n}} \sum_{g \in S} g. \label{eq:rho}
\end{equation}
The reduced density matrix $\rho_A=\Tr_{\bar{A}}\rho$ can be evaluated by taking the partial trace over individual stabilizer group elements.

If the support of $g$ intersects with $\bar{A}$, then $g$ may be expressed as $g=\gamma_1\dots\gamma_m\otimes h$ where $h$ has support exclusively in $A$. Since the first factor has vanishing trace, it follows that $\Tr_{\bar{A}}g=0$. Thus
\begin{equation}
    \rho_A = \frac{1}{2^{2n}} \sum_{g \in S} \text{Tr}_{\bar{A}} g = \frac{1}{2^{2n_A}} \sum_{g \in S_A} g
\end{equation}
where $n_A$ is the number of Majorana modes in $A$ and $S_A$ is the stabilizer subgroup generated by elements $g$ with support exclusively in $A$. This operator is proportional to the projector on to the subspace stabilized by $S_A$, which has dimension $2^{\left(n_A-|S_A|\right)}$ where $|S_A|$ is the number of independent generators of $S_A$. The entanglement entropy is therefore
\begin{equation}
    E_A = - \text{Tr} \rho_A \log \rho_A = n_A-|S_A|.
\end{equation}

The corresponding `doubled' CSS code has $2n$ qubits, $n$ independent $X$ type stabilizer generators, and $n$ independent $Z$ type generators. The entanglement entropy of region $A$ is\cite{entanglementStabilizer}
\begin{equation}
    E^\textrm{CSS}_A = 2n_A-|S^\textrm{CSS}_A|= 2n_A-2|S_A|=2E_A.
\end{equation}

\section{Mapping the Majorana color code to the toric code}
\label{sec:2dmajoranamapping}

In this appendix, we briefly discuss a unitary mapping which decouples the fermionic modes of the Majorana color code on the square-octagon lattice\cite{PhysRevX.5.041038,PhysRevB.97.205404,MFC} from its underlying toric code topological order.

\begin{figure}[htbp]
    \centering
    \includegraphics[width=0.21\textwidth]{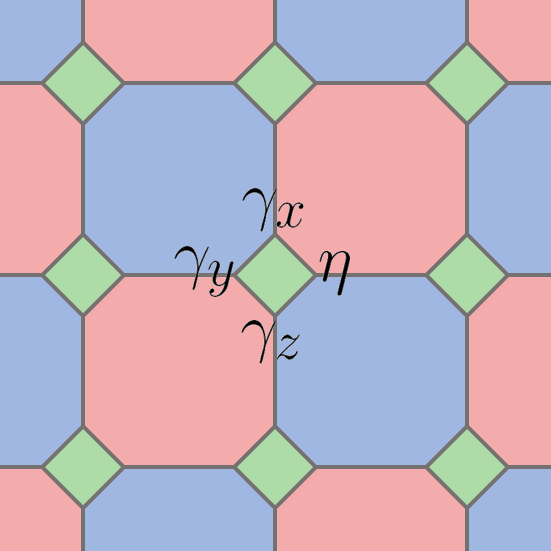}
    \qquad
    \includegraphics[width=0.21\textwidth]{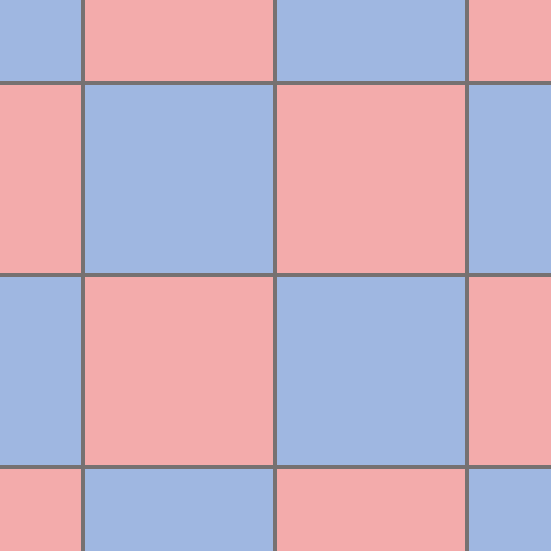}
    \caption{(Left) Square-octagon lattice of the Majorana color code, containing one Majorana fermion at each vertex. The 4 Majoranas around each green plaquette are labelled $\eta$, $\gamma_x$, $\gamma_y$ and $\gamma_z$. (Right) Square lattice of the Wen plaquette model, containing a qubit and two ancillary Majoranas $\gamma_A$ and $\gamma_B$ at each vertex.}
    \label{fig:wenplaquette}
\end{figure}

In this model, one Majorana fermion lies at each vertex of the square-octagon lattice (Fig. \ref{fig:wenplaquette}). The Hamiltonian has the form
\begin{equation}
    H = - \sum_p O_p
\end{equation}
where $p$ runs over all plaquettes, square or octagonal, and $O_p$ takes the form
\begin{equation}
    O_p \equiv \prod_{v \in p} i^{1/2} \gamma_v.
\end{equation}
Since the square-octagon lattice is three-colorable, the plaquette terms are mutually commuting and unfrustrated.

\begin{figure}[htbp]
    \includegraphics[width=0.35\textwidth]{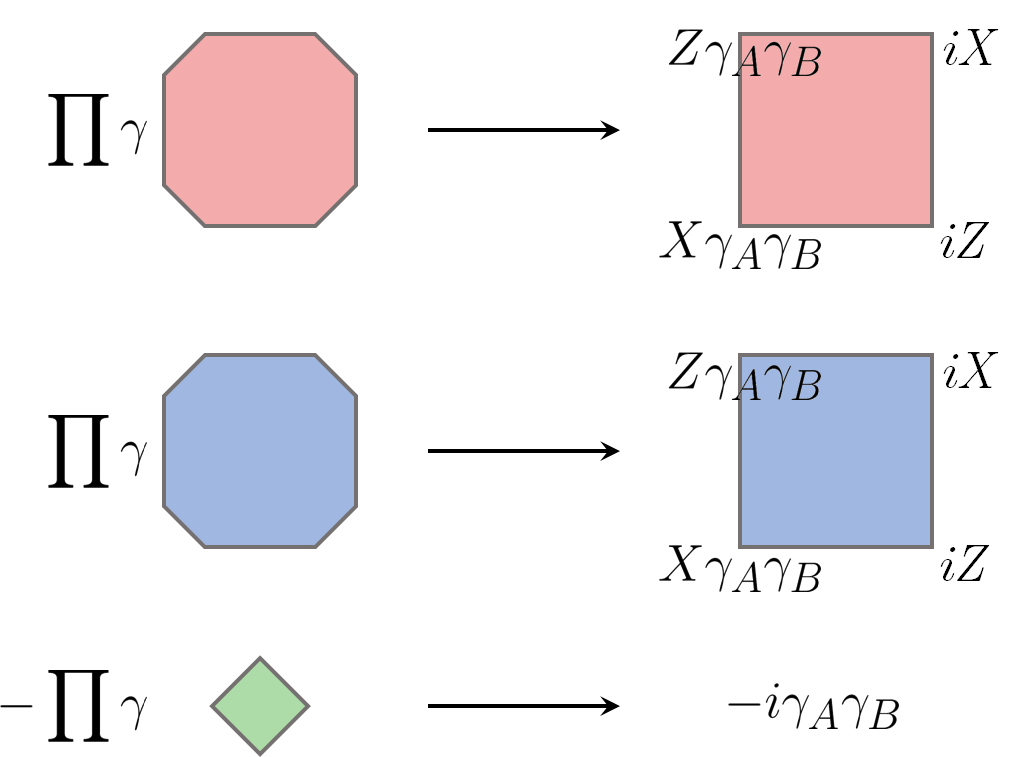}
    \caption{Transformation of plaquette stabilizers under the unitary mapping defined by (\ref{eqn:colorcode}).}
    \label{fig:colorcodemapping}
\end{figure}

To decouple the fermionic modes, we identify the 4 Majorana Hilbert space around each green square plaquette with the Hilbert space of one qubit and 2 Majoranas. Denote the 4 Majoranas by $\eta$, $\gamma_x$, $\gamma_y$, and $\gamma_z$ (as shown in Fig. \ref{fig:colorcodemapping}), and the Pauli operators and 2 Majoranas of the latter space by $X$, $Z$, $\gamma_A$, and $\gamma_B$. We can unitarily map between these two Hilbert spaces according to the following transformation of operators:
\begin{equation}
    \eta\to\gamma_A,\quad \gamma_x\to\gamma_BX,\quad \gamma_y\to\gamma_BY,\quad \gamma_z\to\gamma_BZ
    \label{eqn:colorcode}
\end{equation}
where $Y=iXZ$ is the Pauli operator. This local mapping preserves the commutation relations and the fermionic parity, hence it represents a parity-preserving local unitary operator.

The plaquette terms of the Majorana color code Hamiltonian transform according to Fig. \ref{fig:colorcodemapping}. In particular, the green square terms $-\eta\gamma_x\gamma_y\gamma_z$ are mapped into stabilizer generators for the ancillary fermionic modes, $-i\gamma_A\gamma_B$, whereas the red and blue octagon terms are mapped into stabilizer generators $X_{i}Z_{i+\hat{x}}X_{i+\hat{x}+\hat{y}}Z_{i+\hat{y}}$ of the Wen plaquette model\cite{PhysRevLett.90.016803} (modulo two nearby fermionic stabilizers), which is local unitarily equivalent to the toric code.

\section{Double semion model on a square lattice}
\label{sec:ds}

In this appendix, we briefly discuss a local unitary transformation that allows one to write the double semion model, originally defined on the honeycomb lattice,\cite{Stringnet} as a model of qubits on the edges of a square lattice.

\begin{figure*}[hbtp]
    \centering
    \vspace{.5cm}
    \subfloat[]{
    \includegraphics[width=0.22\textwidth]{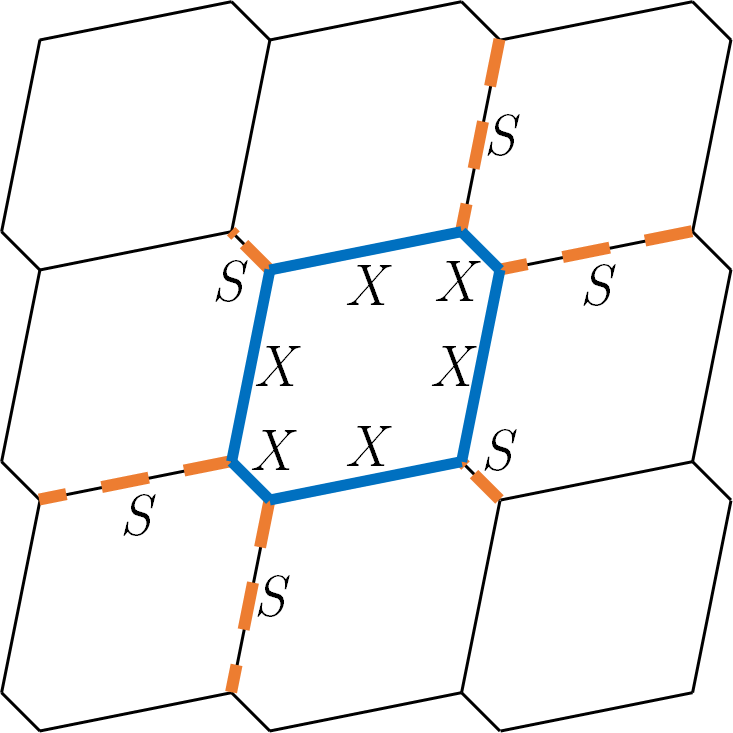}} \qquad
    \subfloat[]{
    \includegraphics[width=0.13\textwidth]{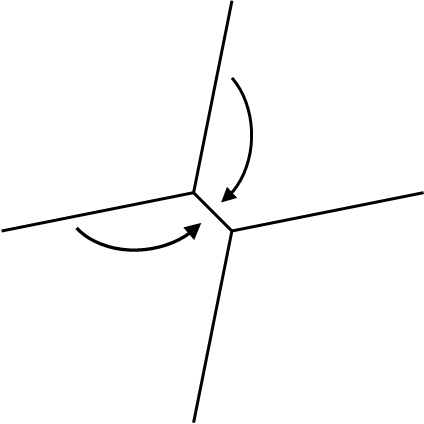}}\qquad
    \subfloat[]{
    \includegraphics[width=0.22\textwidth]{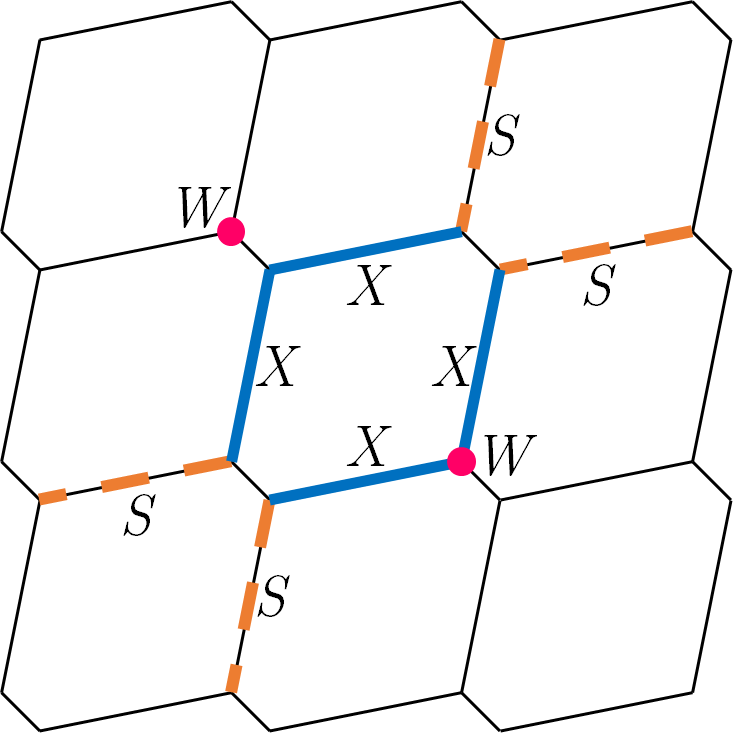}} \qquad
    \subfloat[]{
    \includegraphics[width=0.22\textwidth]{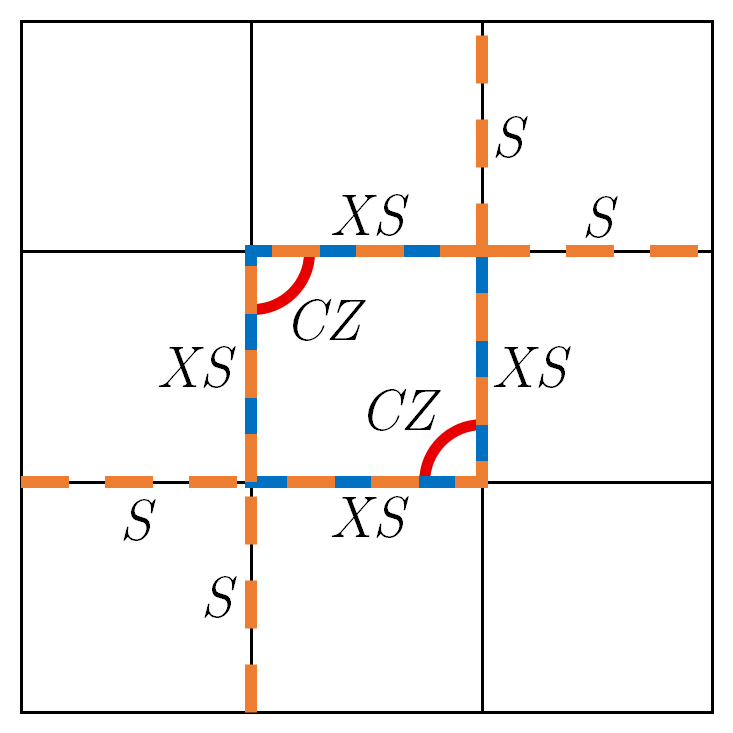}}\\ \vspace{.5cm}
    \subfloat[]{
    \includegraphics[width=0.35\textwidth]{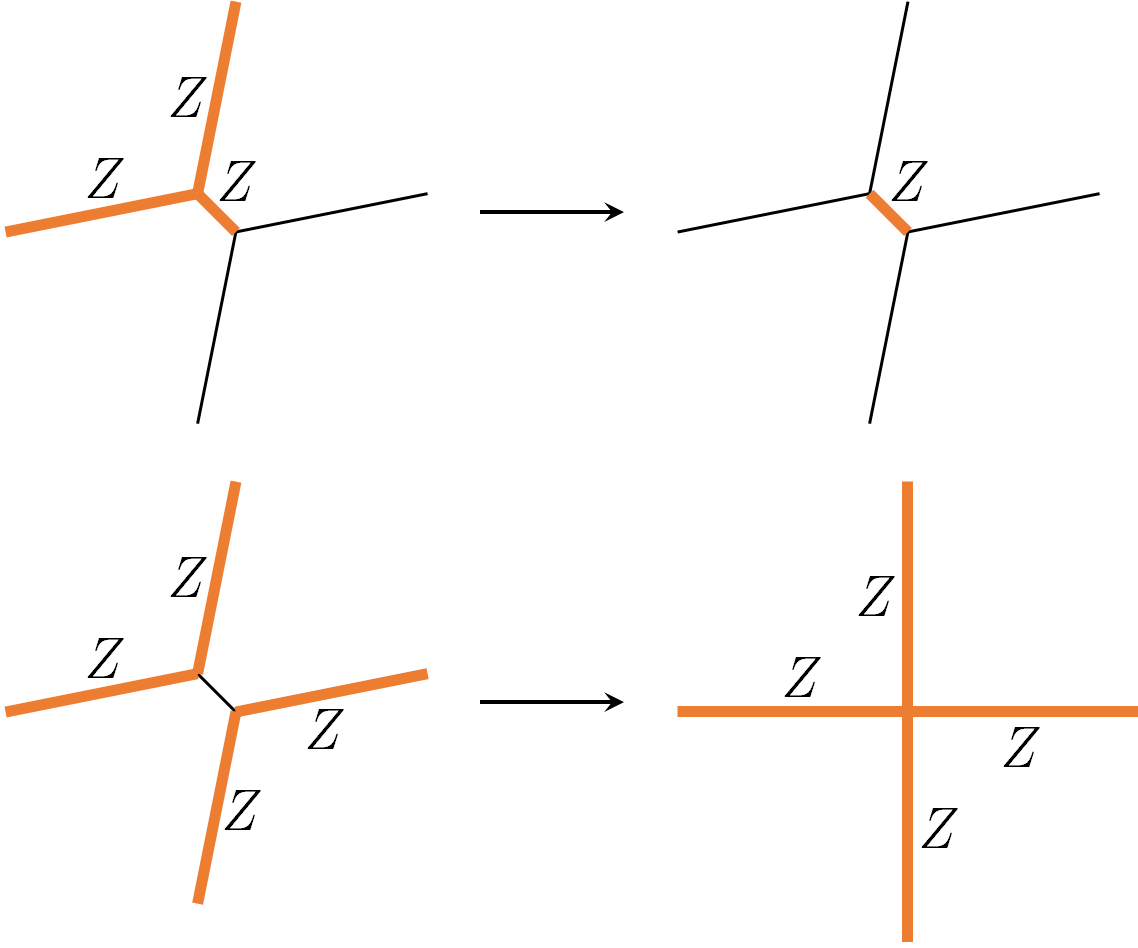}}\qquad
    \subfloat[]{
    \includegraphics[width=0.5\textwidth]{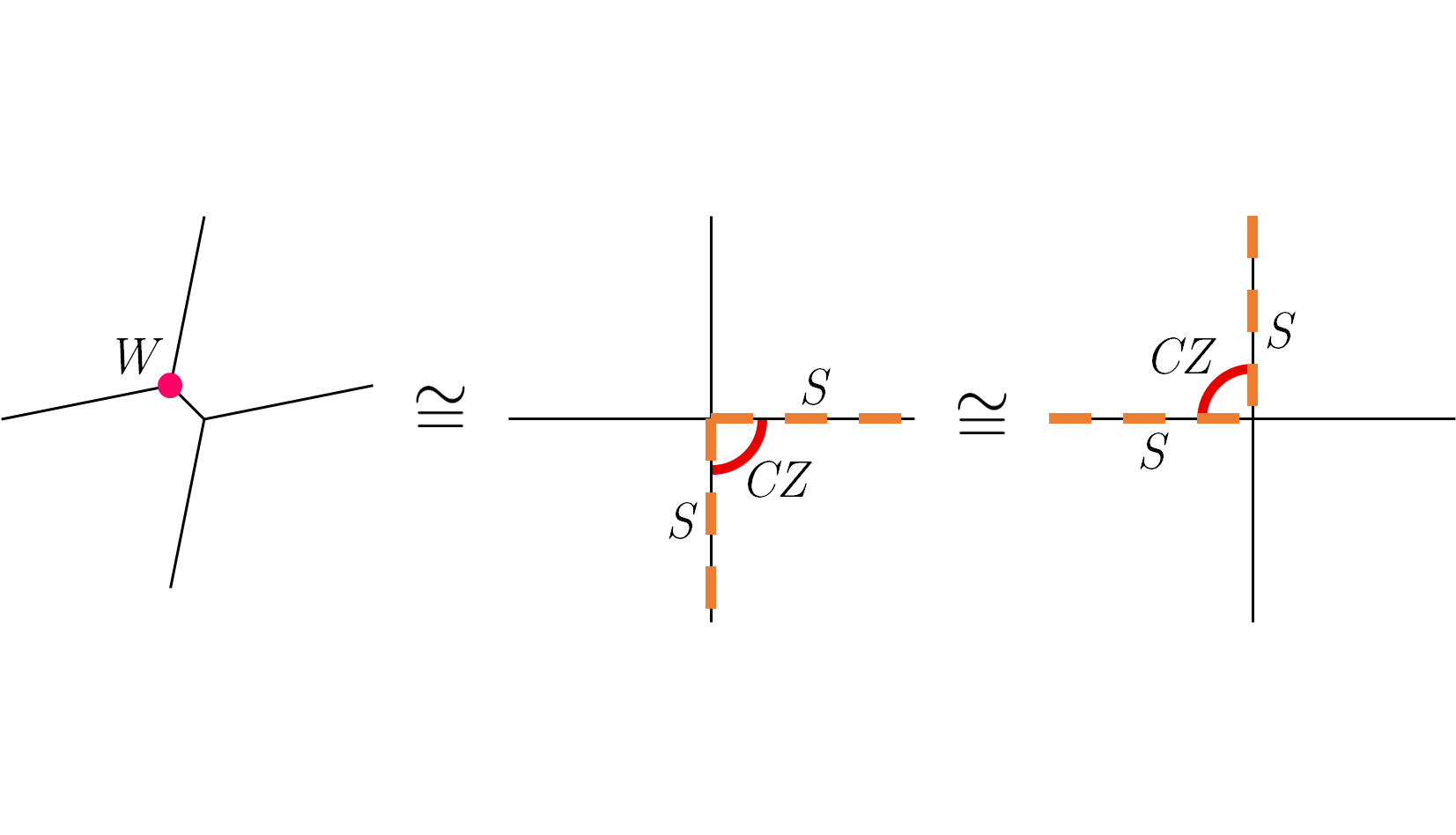}}
    \caption{(a) The component $B_h$ of the double semion plaquette term on the honeycomb lattice. (b) A unit cell of the local unitary $U$ which disentangles the short edge qubits from the rest of the system. The arrow represents the $CX$ gate with control at the tail and target at the head. (c) The image operator $B_h'=U^\dagger B_hU$ (here, the $W$ gates precede the $X$ gates, and act on the 3 edges adjacent to the magenta vertices). (d) The component $B_p$ of the plaquette term on the square lattice (here, the $CZ$ gates precede the $XS$ gates). (e) Mapping of vertex constraints under conjugation by $U$. The top constraints become terms in the ancillary Hamiltonian $H_0$, whereas the bottom constraints (product of two vertex constraints) become the vertex terms of the square lattice Hamiltonian. (f) Operator relations which hold within the subspace satisfying the vertex (and ancillary) constraints.}
    \label{fig:DSHamiltonian}
\end{figure*}

The double semion model contains one qubit on each edge of the honeycomb lattice, and has Hamiltonian
\begin{equation}
    H=-\sum_v A_v-\sum_p\tilde{B}_h
\end{equation}
where $v$ indexes vertices and $h$ indexes hexagonal plaquettes. The vertex constraint is $A_v=Z_1Z_2Z_3$ acting on the 3 adjacent edges, and the hexagon term is \begin{equation}
\begin{split}
    \tilde{B}_h=B_h\prod_{v\in p}\frac{1+A_v}{2}, \\
    B_h=\prod_{e\in h}X_e\prod_{l\in h}S_l.
\end{split}
\end{equation}
Here $e$ runs over the 6 edges of hexagon $h$, whereas $l$ runs over the 6 legs external to $h$, as shown in Fig. \ref{fig:DSHamiltonian}(a). $S=i^{\frac{1-Z}{2}}$ is the phase gate.

It is possible to disentangle the qubits lying on the short edges of the honeycomb lattice from the rest of the system, leaving behind a square lattice. In particular, the unitary operator $U$ accomplishes this task, which is a translation-invariant array of $CX$ gates as shown in Fig. \ref{fig:DSHamiltonian}(b). To be precise,
\begin{equation}
    U^\dagger H U\cong H' + H_0
\end{equation}
where $H_0$ stabilizes the ancillary qubits and
$H'$ is the double semion Hamiltonian on the square lattice:
\begin{equation}
    H'=-\sum_vA_v-\sum_p\tilde{B}_p.
\end{equation}
Here $A_v=Z_1Z_2Z_3Z_4$, acting on the 4 adjacent edges, and
\begin{equation}
    \tilde{B}_p=B_p\prod_{v\in p}\frac{1+A_v}{2}, \\
\end{equation}
where $B_p$ is depicted graphically in Fig. \ref{fig:DSHamiltonian}(d). The relation $\cong$ indicates that the two sides have identical ground spaces.

To see this, note that $S_3\to W_{123}$ under conjugation by $CX_{13}CX_{23}$, where we have defined $W_{123}=i^\frac{1-Z_1Z_2Z_3}{2}$, and thus $B_h$ is mapped to the operator $B_h'$ shown in Fig. \ref{fig:DSHamiltonian}(c). Moreover, $U$ maps the original vertex constraints according to Fig. \ref{fig:DSHamiltonian}(e), yielding the vertex terms on the square lattice as well as the ancillary terms comprising $H_0$. Finally, $B_h'$ is equivalent to $B_p$ in the subspace satisfying the vertex constraints, due to the relations shown in Fig. \ref{fig:DSHamiltonian}(f).

\end{document}